\documentclass[lettersize,journal]{IEEEtran}
\usepackage{amsmath,amsfonts}
\usepackage{algorithmic}
\usepackage{algorithm}
\usepackage{array}
\usepackage[caption=false,font=normalsize,labelfont=sf,textfont=sf]{subfig}
\usepackage{textcomp}
\usepackage{stfloats}
\usepackage{url}
\usepackage{verbatim}
\usepackage{graphicx}
\usepackage[dvipsnames]{xcolor}
\usepackage{cite}
\DeclareGraphicsExtensions{.pdf,.png,.jpg}

\newcommand {\Define} {\stackrel {\Delta} {=}  }
\newcommand{\mya}{\mathrel{\overset{\makebox[0pt]{{\tiny(a)}}}{=}}}
\newcommand{\myb}{\mathrel{\overset{\makebox[0pt]{{\tiny(b)}}}{=}}}
\newcommand{\myc}{\mathrel{\overset{\makebox[0pt]{{\tiny(c)}}}{=}}}
\newcommand{\myd}{\mathrel{\overset{\makebox[0pt]{{\tiny(d)}}}{=}}}

\newtheorem{theorem}{Theorem}

\usepackage{cite}
\usepackage{comment}

\begin{document}
	\title{Zak-OTFS based Multiuser Uplink in Doubly-Spread Channels}
	\author{\IEEEauthorblockN{Imran Ali Khan, Saif Khan Mohammed, Ronny Hadani, Ananthanarayanan Chockalingam and Robert Calderbank, Fellow, IEEE}
		\IEEEauthorblockA{ \thanks{}}
	}
	\maketitle
	
	\begin{abstract}
    Wireless users with different characteristics will be expected to share spectrum in next generation communication networks. One of the great strengths of wireless networks based on Orthogonal Frequency Division Multiplexing (OFDM) is the ease with which different non-overlapping time-frequency (TF) resources can be allocated to different users by simply shifting each user’s signal in time and frequency. However, a significant weaknesses of OFDM is the inflexibility of sub-carrier spacing. Since OFDM does not allow users to have different sub-carrier spacing, a single user subject to inter-carrier interference causes carrier spacing to increase for all users. Zak-OTFS is an alternative delay-Doppler (DD) domain modulation scheme, where, in contrast to OFDM, the Input-Output (I/O) relation is predictable. We match the strength of OFDM by designing a novel DD domain method of shaping the transmitted Zak-OTFS pulse on the uplink that enables flexible non-overlapping TF resource allocation. The base station (BS) receives a superposition of uplink signals and applies individual matched filters to obtain the data specific to individual users. We develop theoretical measures of interference between users, and present numerical simulations for a vehicular channel model representative of next generation propagation environments. We demonstrate single-user performance in a multiuser Zak-OTFS uplink system without needing to provision guard bands between TF resources allocated to different users. These performance results demonstrate that the benefits of a predictable Zak-OTFS waveform can be realized within an architecture for uplink communication.
    \end{abstract} 
	
	\begin{IEEEkeywords}
		ZAK-OTFS, multiuser, uplink, Doubly spread.
	\end{IEEEkeywords}

\normalsize
\section{Introduction}
Next generation communication networks are expected to support
diverse communication scenarios where the channel delay and Doppler spread can be several tens of micro seconds and several KHz. Some  examples include non-terrestrial networks (NTN), Unmanned Aerial Vehicle (UAV) and aircraft communication, and high speed train \cite{IMT2030}. Integration of
terrestrial networks and NTN is being actively considered to provide
seamless and ubiquitous connectivity \cite{NTN}.
Multiuser uplink communication in current cellular wireless networks are based on Orthogonal Frequency Division Multiple Access (OFDMA)
which is known to suffer from severe inter-carrier interference (ICI)
in scenarios characterized by high Doppler spread. In order to
mitigate the effect of ICI, the OFDM sub-carrier spacing is chosen sufficiently larger than the largest possible channel Doppler shift.
In a multiuser OFDMA uplink, the system carrier spacing is therefore decided by the user which experiences the largest Doppler shift. Due to this non-flexible numerology in OFDM, all the other users (majority of whom might not be experiencing high Doppler) would also have to use a large carrier spacing which mean loss in effective throughput
due to a significant fraction of time resource being used by the cyclic prefix (CP) whose duration is determined by the channel delay spread and cannot be made arbitrarily small.

Multi-carrier (MC) OTFS modulation proposed in \cite{Hadani2017, embedded}
has been shown to be robust to channel induced delay and Doppler spread and achieve significantly better performance than OFDM. In MC-OTFS, information is embedded in the delay-Doppler (DD) domain and channel estimation and acquisition are also carried out in the DD domain. There are several existing analyses of MC-OTFS based multiuser uplink \cite{MCOTFSMA1, MCOTFSMA2, MCOTFSMA4, MCOTFSMA5, MCOTFSMA6, MCOTFSMA7}. Although MC-OTFS performs better than OFDM, its performance degrades at high Doppler spreads \cite{twosteppaper}. This is because the MC-OTFS input-output (I/O) relation is not predictable, i.e., the channel response to an arbitrary MC-OTFS carrier cannot be predicted/estimated accurately from the response to a particular MC-OTFS carrier (e.g., a pilot carrier). Therefore, acquiring the MC-OTFS I/O relation with low-overhead is challenging and inaccurate estimates result in performance degradation for high Doppler spreads \cite{otfsbook}. 

Zak transform based DD domain receiver signal processing was proposed
in \cite{twosteppaper,derivationpaper} and shown to achieve better robustness towards channel Doppler spread when compared to receiver processing in MC-OTFS.
Zak transform based OTFS modulation (Zak-OTFS) was introduced in \cite{ZAKOTFS1, ZAKOTFS2, otfsbook} and
is different from MC-OTFS. In Zak-OTFS, information symbols are carried by narrow pulses in the DD domain which are quasi-periodic functions
having period $\tau_p$ and $\nu_p = 1/\tau_p$ along the delay and Doppler axis respectively. The Zak-OTFS I/O relation is predictable
when the crystallization condition is satisfied, i.e., the channel delay and Doppler spread are less than the delay and Doppler period respectively. Therefore, a single DD pilot carrier is sufficient to acquire the Zak-OTFS I/O relation resulting in low channel acquisition overhead when compared to MC-OTFS \cite{spreadpaper,twobytwopaper}. Even with a single pilot carrier, the acquired channel estimates are accurate and therefore the performance is almost invariant of the Doppler spread as long as the crystallization condition is satisfied.

There has been some work on multiuser random-access in Zak-OTFS \cite{Sandesh25}. However, there is
no work on Zak-OTFS multiuser uplink (Zak-OTFS-MUL) where
users request and are allocated/granted \emph{dedicated} time-frequency (TF) resource. In this paper, we consider a Zak-OTFS-MUL system where users are allocated dedicated non-overlapping TF resources.
We propose novel DD domain pulse shaping of the transmitted Zak-OTFS signal in the uplink which allows for
flexible TF resource allocation, i.e., a user can be allocated resource
restricted to any arbitrary TF region.
This flexible nature of the proposed Zak-OTFS-MUL system makes it compatible with the 3GPP 5G NR numerology, i.e., we can choose the time and bandwidth allocated to a user to be exactly same as those allocated in 3GPP 5G NR.
While OFDMA in not flexible in the choice of sub-carrier spacing, i.e., all users need to transmit OFDM sub-carriers with the same sub-carrier spacing which leads to unnecessarily longer CP than required, with Zak-OTFS-MUL we can choose the delay and Doppler period of each user independently of other users so as to optimize that user's performance.
For users with high Doppler spread (e.g., high mobility scenarios) we choose a large Doppler period and for those experiencing small Doppler spread (e.g., low mobility scenarios) we can choose a smaller Doppler period.
This flexibility allows Zak-OTFS-MUL to support the integration of different types of networks (e.g., terrestrial networks and NTN) to provide seamless and ubiquitous connectivity.

In Section \ref{secsystemmodel} we present the Zak-OTFS signal processing at each user transmitter and at the BS receiver. Each user uses its own transmit pulse shaping DD filter.
The BS receives the superposition of the Zak-OTFS signals transmitted from all uplink users, each user's signal experiencing a different delay-Doppler channel between itself and the BS.
At the BS receiver, to detect a particular user's information symbols, we match-filter the received signal with a DD domain filter which is matched to the transmit pulse-shaping DD filter of that user.
For each user, we derive the relation between the match-filtered output for that user and the DD domain information signals transmitted by all users. The match-filtered output of each user is used to detect its information symbols, by treating the multiuser interference (MUI) in match-filtered output as noise. This allows us to carry out the detection of each user's information symbols separately, instead of a more complex joint multiuser detection.

In Section \ref{secproposedMA}, we propose a novel pulse shaping DD filter which allows for flexible non-overlapping TF resource allocation in Zak-OTFS-MUL.
The resource allocation is indeed non-overlapping for an ideal AWGN channel without delay or Doppler spread, i.e., the ratio of interference energy received from a user in the match-filtered output of some other user to the useful signal energy received in the match-filtered output of the user which transmitted the signal, is much smaller than one (see Section \ref{tflocal}).
We subsequently refer to this ratio as the interference  to useful signal ratio. In Section \ref{embedpilot}, we discuss the structure of a Zak-OTFS frame/packet transmitted by a user in the uplink, i.e., the DD location of carriers assigned for data and those assigned for the single pilot and the surrounding guard region. 

In Section \ref{numsec} we present numerical simulations of the bit error rate (BER) and normalized mean square channel estimation error (NMSE) performance of the proposed Zak-OTFS-MUL system for a vehicular-A channel model \cite{EVAITU}.
With a sinc pulse shaping filter, we observe that the interference to useful signal ratio is smaller than $-30$ dB even for a high Doppler spread of $6$ KHz and a delay spread of $2.5 \mu s$. Since the signal to noise ratio in real world systems is usually less than $30$ dB, the
BER and NMSE performance is limited by AWGN and not by MUI.
We therefore observe single-user performance in a multiuser Zak-OTFS uplink system. In the proposed Zak-OTFS-MUL system we do not provision for guard TF regions separating adjacent TF allocations to different users. Yet, we achieve single-user performance. This is because, firstly the interference is dominated by AWGN and secondly, even though the MUI is restricted to the boundary between adjacent TF allocations, in the DD domain it is spread almost uniformly across all DD carriers. This happens because even though a DD carrier is a narrow pulse in the DD domain, its energy is spread almost uniformly across the allocated TF region. Single-user performance in multiuser uplink shows that the benefits of a predictable Zak-OTFS waveform can be realized within an architecture for uplink communication that enables users with different delay-Doppler characteristics to share spectrum without the need for guard resource separating their TF allocations.

     \begin{figure*}
    \centering
          \includegraphics[scale=0.35]{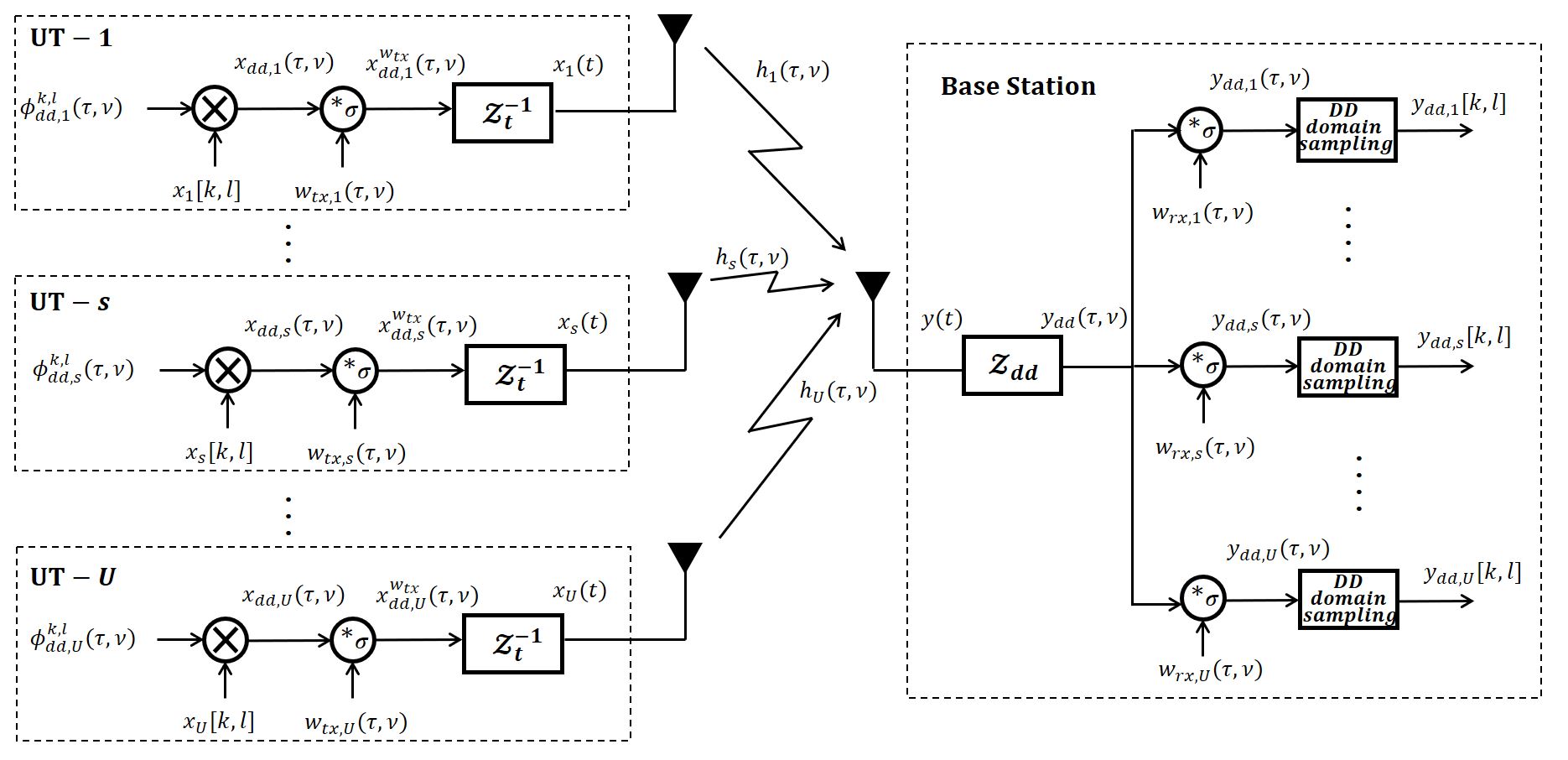}
			\caption{Transmitter and Receiver processing in Zak-OTFS-based uplink multiuser system}
			\label{fig:system_model}
    \end{figure*}
\section{System model} 	
\label{secsystemmodel}
    This paper considers a Zak-OTFS based uplink communication system as illustrated in Fig. \ref{fig:system_model}, where $U$ single antenna user terminals (UTs) communicate with a single antenna base station (BS).
    
    In a Zak-OTFS based system, information symbols of the $s$-th UT ($s=1,2,\cdots, U$) are carried by carriers which are quasi-periodic pulses in the two-dimensional DD domain, localized at points of the information grid/lattice $\Lambda_{s}$ given by \cite{ZAKOTFS2, otfsbook}

    {\small
    \vspace{-4mm}
        \begin{eqnarray}
        \Lambda_s & \Define & \left\{  \left( \frac{k}{B_s} \,,\, \frac{l}{T_s} \right)  \, {\Big \vert} \, k, l \in {\mathbb Z} \right\}.
    \end{eqnarray}\normalsize}Here, $T_s$ and $B_s$ respectively denote the approximate time-duration and bandwidth
    of the uplink signal transmitted by the $s$-th UT.
    Let $\tau_{p,s}$ and $\nu_{p,s} \Define 1/\tau_{p,s}$ respectively denote the delay and Doppler period of the DD domain carrier pulses carrying the information symbols of the $s$-th UT. Let $M_s \Define B_s \tau_{p,s}$ and $N_s \Define T_s \nu_{p,s}$ be integers.
    The $s$-th UT transmits $M_s \times N_s$ information symbols $x_s[k,l]$, $k=0, 1, \cdots, M_s-1$ and $l=0, 1, \cdots, N_s-1$.
    
    The $(k,l)$-th information symbol $x_s[k,l]$ is carried by the DD function
    $\psi^{k,l}_{dd,s} (\tau, \nu)$ which is a quasi-periodic DD function, it is periodic along the Doppler axis with period
    $\nu_{p,s}$ and is quasi-periodic along the delay axis with period $\tau_{p,s}= 1/\nu_{p,s}$, i.e.
    for all $n,m \in {\mathbb Z}$

    {\small
    \vspace{-4mm}
    \begin{eqnarray}
    \label{eqn1}
     \psi^{k,l}_{dd,s} (\tau + n \tau_{p,s}, \nu + m \nu_{p,s}) & \hspace{-1.5mm} = & \hspace{-1.5mm} e^{j 2 \pi n \nu \tau_{p,s}} \, \psi^{k,l}_{dd,s} (\tau , \nu ).
    \end{eqnarray}\normalsize}This function is obtained by DD domain filtering (twisted convolution \cite{otfsbook}) of another DD domain quasi-periodic function $\phi^{k,l}_{dd,s}(\tau, \nu)$ with the pulse-shaping filter $w_{tx,s}(\tau, \nu)$, i.e.

    {\small
    \vspace{-4mm}
    \begin{eqnarray}
    \label{eqn2}
        \psi^{k,l}_{dd,s} (\tau , \nu ) & = & w_{tx,s}(\tau, \nu) \, *_{\sigma}  \, \phi^{k,l}_{dd,s}(\tau, \nu) \nonumber \\
        & & \hspace{-21mm} = \hspace{-1mm} \iint \hspace{-1mm}  w_{tx,s}(\tau', \nu') \, \phi^{k,l}_{dd,s}(\tau - \tau', \nu - \nu') \, e^{j 2 \pi \nu' (\tau - \tau')} \, d\tau' d\nu',
    \end{eqnarray}\normalsize}where $*_{\sigma}$ denotes the twisted convolution operation and $\phi^{k,l}_{dd,s}(\tau, \nu)$ is a quasi-periodic Dirac-delta DD pulse at the DD point $\left( \frac{k \tau_{p,s}}{M_s} \,,\, \frac{l \nu_{p,s}}{N_s}\right)$, i.e.
        
        {\vspace{-4mm}
        \small
    \begin{eqnarray}
        \label{basis_sig}
        \phi^{k,l}_{dd,s} (\tau, \nu) &\hspace{-1mm}=&\hspace{-1mm}  \sum_{n,m \in {\mathbb Z}} \Bigg [ e^{j 2 \pi \frac{nl}{N_s}} \delta\hspace{-1mm}\left(\hspace{-1mm}\tau - n\tau_{p,s} - \frac{k \tau_{p,s}}{M_s} \hspace{-1mm}\right) \nonumber \\ 
        && \hspace{2.2cm}\delta\hspace{-1mm}\left(\hspace{-1mm}\nu - m\nu_{p,s} - \frac{l \nu_{p,s}}{N_s}\hspace{-1mm}\right) \Bigg ].
    \end{eqnarray}
    \normalsize}
    The DD domain information signal carrying all $M_s \times N_s$ symbols is given by the linear sum
    \begin{eqnarray}
    \label{eqn5}
        x_{dd,s}^{w_{tx}}(\tau, \nu) & = & \sum\limits_{k=0}^{M_s - 1} \sum\limits_{l=0}^{N_s -1} x_s[k,l] \, \psi^{k,l}_{dd,s} (\tau , \nu ) \nonumber \\
        & \mya & w_{tx,s}(\tau, \nu) \, *_{\sigma}  \,  x_{dd,s}(\tau, \nu),  \nonumber \\
        x_{dd,s}(\tau, \nu) & \Define & \sum\limits_{k=0}^{M_s - 1} \sum\limits_{l=0}^{N_s -1} x_s[k,l] \, \phi^{k,l}_{dd,s} (\tau , \nu )
    \end{eqnarray}where step (a) follows from the linearity of twisted convolution. The TD signal transmitted by the $s$-th UT is given by the inverse Zak-transform (see \cite{ZAKOTFS1, ZAKOTFS2, otfsbook}) of $x_{dd,s}^{w_{tx}}(\tau, \nu)$, i.e.
    \begin{eqnarray}
        x_s(t) & = & {\mathcal Z_t}^{-1}\left( x_{dd,s}^{w_{tx}}(\tau, \nu) \right) \nonumber \\
        & = & \sqrt{\tau_{p,s}} \int\limits_{0}^{\nu_{p,s}} x_{dd,s}^{w_{tx}}(t, \nu) \, d\nu.
    \end{eqnarray}Hence, $x_{dd,s}^{w_{tx}}(\tau, \nu)$ is the DD domain representation of $x_s(t)$ given by its Zak-transform, i.e.
    \begin{eqnarray}
    \label{eqn7}
        x_{dd,s}^{w_{tx}}(\tau, \nu) & = & {\mathcal Z}_t \left( x_s(t) \right) \nonumber \\
        & = & \sqrt{\tau_{p,s}} \sum\limits_{k \in {\mathbb Z}} x_s(\tau + k \tau_{p,s}) \, e^{-j 2 \pi k \nu \tau_{p,s}}.
    \end{eqnarray}The DD realization of any TD signal is a quasi-periodic function (see (\ref{eqn1})) and this is why inverse Zak-transform is only defined for quasi-periodic DD functions. For any
    \begin{eqnarray}
        \label{eqn8p1}
        a_{dd}(\tau, \nu) & =  & w(\tau, \nu) \, *_{\sigma} \, b_{dd}(\tau, \nu),
    \end{eqnarray}it is known that a quasi-periodic $b_{dd}(\tau, \nu)$ results in a quasi-periodic $a_{dd}(\tau, \nu)$ (since twisted convolution preserves quasi-periodicity, see \cite{otfsbook}) and also that the TD realization $a(t)$ and $b(t)$ of $a_{dd}(\tau, \nu)$ and $b_{dd}(\tau, \nu)$ respectively, are related by
    \begin{eqnarray}
    \label{eqn8}
        a(t) & = & \iint w(\tau, \nu) b(t - \tau) \, e^{j 2 \pi \nu (t - \tau)} \, d\tau \, d\nu.
    \end{eqnarray}For sake of completeness, we have provided a proof of (\ref{eqn8}) in Appendix \ref{appprfb} (this result has been taken from \cite{ZakGauss}). Using this result in $x_{dd,s}^{w_{tx}}(\tau, \nu) = w_{tx,s}(\tau, \nu) \, *_{\sigma}  \,  x_{dd,s}(\tau, \nu)$ gives
    \begin{eqnarray}
    \label{eqn9}
        x_s(t) & = & \iint w_{tx}(\tau, \nu) x_{\delta, s}(t - \tau) \, e^{j 2 \pi \nu (t - \tau)} \, d\tau \, d\nu
    \end{eqnarray}where $x_{\delta, s}(t)$ is the TD realization (inverse Zak-transform)
    of $x_{dd,s}(\tau, \nu)$ and due to the linearity of the inverse Zak-transform it is given by
    \begin{eqnarray}
        x_{\delta, s}(t) & = & \sum\limits_{k=0}^{M_s -1} \sum\limits_{l=0}^{N_s -1 } x_s[k,l] \, \phi^{k,l}_{s}(t),
    \end{eqnarray}where $\phi^{k,l}_{s}(t)$ is the inverse Zak-transform of $\phi^{k,l}_{dd,s}(\tau, \nu)$ which is given by (see \cite{otfsbook})
    \begin{eqnarray}
    \label{eqn1213}
        \phi^{k,l}_{s}(t) & = & {\mathcal Z}_t^{-1} \left( \phi^{k,l}_{dd,s}(\tau, \nu) \right) \nonumber \\
        &  &  \hspace{-16mm} \mya  \sqrt{\tau_{p,s}} \sum_{n \in \mathbb{Z}} e^{j 2 \pi \frac{nl}{N_s}} \delta \left( t - n \tau_{p,s} -\frac{k\tau_{p,s}}{M_s}\right) \nonumber \\
        & & \hspace{-16mm} = \sqrt{\tau_{p,s}} e^{j 2 \pi \frac{l (t - \tau_{p,s})}{N_s \tau_{p,s}} } \, \sum_{n \in \mathbb{Z}} \delta \left( t - n \tau_{p,s} -\frac{k\tau_{p,s}}{M_s}\right)
    \end{eqnarray}where step (a) follows from Appendix \ref{appendix_td_pulsone} (reproduced here from \cite{otfsbook} for the sake of completeness). Note that $\phi^{k,l}_{s}(t)$ is a Dirac-delta pulse train of infinite duration (with pulses at the delay locations of $\phi^{k,l}_{dd,s}(\tau, \nu)$) multiplied (modulated) with a sinusoid (tone) of frequency $l/T_s = l/(N_s \tau_{p,s})$, and is therefore appropriately called as a Dirac-delta \emph{pulsone}. Note that these Dirac-delta pulsones have infinite bandwidth and time-duration and therefore $x_{\delta, s}(t)$ also has infinite time duration and bandwidth. Next, we see that pulse shaping of $x_{\delta, s}(t)$ with an appropriate DD filter $w_{tx}(\tau, \nu)$ as in (\ref{eqn9}) results in the TD transmit signal $x_s(t)$ which is approximately time and bandwidth-limited to $T_s$ seconds and $B_s$ Hz.
    
    Taking the inverse Zak-transform of both sides of (\ref{eqn5}) gives
    \begin{eqnarray}
    \label{eqn132}
        x_s(t) & = & \sum\limits_{k=0}^{M_s -1} \sum\limits_{l=0}^{N_s -1 } x_s[k,l] \, \psi^{k,l}_s(t),
    \end{eqnarray}where $\psi^{k,l}_s(t)$ carries the $(k,l)$-th information symbol $x_s[k,l]$ and is the inverse Zak-transform (i.e, TD realization) of $\psi^{k,l}_{dd,s}(\tau, \nu)$.
    From (\ref{eqn2}) it follows that
    \begin{eqnarray}
        \psi^{k,l}_{dd,s}(\tau, \nu) & = & w_{tx,s}(\tau, \nu) \, *_{\sigma} \, \phi^{k,l}_{dd,s}(\tau, \nu). 
    \end{eqnarray}Using (\ref{eqn8}) with $a(\tau, \nu) = \psi^{k,l}_{dd,s}(\tau, \nu)$, $w(\tau, \nu) = w_{tx,s}(\tau, \nu)$ and $b(\tau, \nu) = \phi^{k,l}_{dd,s}(\tau, \nu)$ then gives
    \begin{eqnarray}
    \label{eqn14}
        \psi^{k,l}_s(t) & \hspace{-2.5mm} = & \hspace{-2.5mm} \iint w_{tx,s}(\tau, \nu) \, \phi^{k,l}_s(t - \tau) \, e^{j 2 \pi \nu (t - \tau)} \, d\tau \, d\nu.
    \end{eqnarray}With a factorizable pulse shaping filter $w_{tx_s}(\tau, \nu) = w_{B_s}(\tau) \, w_{T_s}(\nu)$, (\ref{eqn14}) gives
    \begin{eqnarray}
    \label{eqn15}
         \psi^{k,l}_s(t) & \hspace{-2mm} = & \hspace{-2mm} w_{B_s}(t) \, \star \, \left[ W_{T_s}(t) \,  \phi^{k,l}_s(t) \right], \,\, \mbox{\small{where}} \nonumber \\
W_{T_s}(t)  & \Define &  \int w_{T_s}(\nu) \, e^{j 2 \pi \nu t} \, d\nu
    \end{eqnarray}is the inverse Fourier transform of $w_{T_s}(\nu)$ (see \cite{ZakGauss} for this result, a derivation has been provided in Appendix \ref{appendix_eqn16} here for the sake of completeness). Fourier transform of both sides of (\ref{eqn15}) gives

    {\vspace{-4mm}
    \small
    \begin{eqnarray}
    \label{eqn187635}
        {\Psi}^{k,l}_{s}(f) & = & \int \psi^{k,l}_s(t) \, e^{-j 2 \pi f t} \, dt, \nonumber \\
        & & \hspace{-10mm} = W_{B_s}(f) \, \left[  w_{T_s}(f) \, \star \, \Phi_s^{k,l}(f) \right],
    \end{eqnarray}\normalsize}where $\Phi_s^{k,l}(f)$ is the Fourier transform of $\phi^{k,l}_s(t)$ and

    {\vspace{-4mm}
\small
\begin{eqnarray}
\label{eqn26}
W_{B_s}(f) & \Define & \int w_{B_s}(\tau) \, e^{-j 2 \pi f \tau} \, d\tau.
\end{eqnarray}\normalsize}
The pulse shaping in (\ref{eqn15}) therefore first limits the Dirac-delta pulsone $\phi^{k,l}_s(t)$ in time through the product with $W_{T_s}(t)$ and then the linear convolution with $w_{B_s}(t)$ limits it in bandwidth.
    We discuss this in greater detail and rigor in Section \ref{subsecDDprocessing} of this paper. As we will show in Section \ref{subsecDDprocessing}, appropriate modifications to the pulse shaping filter will allow us to shift the signal of each UT to non-overlapping regions in the TF domain. 
    
   Let $h_s(\tau, \nu)$ denote the DD spreading function for the physical wireless  channel between the BS and the $s$-th UT. The signal $y(t)$ received at the BS is the sum of signals received from all $U$ UTs \cite{Bello}

    {\vspace{-2mm}
    \small
    \begin{eqnarray}
        \label{rx_sig_t}
        y(t) = \sum_{s=1}^{U} \int \hspace{-2mm} \int h_{s}(\tau, \nu) \, x_{s}(t-\tau) \, e^{j 2 \pi \nu(t-\tau)} d\tau d\nu \, + \, n(t),
    \end{eqnarray}
    \normalsize}where $n(t)$ is AWGN signal with power spectral density $N_{0}$.
    At the BS receiver, Zak transform of $y(t)$ gives its DD representation $y_{dd}(\tau, \nu)$. Zak transform of both sides of (\ref{rx_sig_t}) and using the equivalence between (\ref{eqn8p1}) and (\ref{eqn8}) gives

    {\vspace{-2.5mm}
    \small
    \begin{eqnarray}
        \label{rx_sig}
        y_{dd}(\tau, \nu) = \sum_{s=1}^{U} h_{s}(\tau, \nu) *_{\sigma} x^{w_{tx}}_{dd,s}(\tau, \nu) + n_{dd} (\tau, \nu),
    \end{eqnarray}
    \normalsize}where $n_{dd}(\tau, \nu)$ is the Zak-transform of $n(t)$. At the BS receiver, for detecting the information transmitted by the $q$-th UT, the received DD signal is match-filtered with the filter $w_{rx,q}(\tau, \nu)$ resulting in the match-filtered signal
    \begin{eqnarray}
\label{rxmatchfilt}
        y_{dd,q}(\tau, \nu) & = & w_{rx,q}(\tau, \nu) \, *_{\sigma} \, y_{dd}(\tau, \nu) \nonumber \\
        & & \hspace{-21mm} \mya \sum\limits_{s=1}^U \hspace{-1mm} w_{rx,q}(\tau, \nu) \, *_{\sigma}  h_s(\tau, \nu)  *_{\sigma} \, x^{w_{tx}}_{dd,s}(\tau, \nu) \, +  \, n_{dd,q} (\tau, \nu), \nonumber \\
        n_{dd,q} (\tau, \nu) & \Define & w_{rx,q}(\tau, \nu)  \, *_{\sigma} \, n_{dd}(\tau, \nu),
    \end{eqnarray}where step (a) follows from the expression of $y_{dd,q}(\tau, \nu)$ in the R.H.S. of (\ref{rx_sig}). Further, using $ x^{w_{tx}}_{dd,s}(\tau, \nu) = w_{tx,s}(\tau, \nu) \, *_{\sigma}  \,  x_{dd,s}(\tau, \nu)$ from (\ref{eqn5}), and the fact that twisted convolution is an associative operation (i.e., $a(\tau, \nu) *_{\sigma} (b(\tau, \nu) *_{\sigma} c(\tau, \nu)) = (a(\tau, \nu) *_{\sigma} b(\tau, \nu)) *_{\sigma} c(\tau, \nu)$) it follows that
    \begin{eqnarray}
\label{eqn2020}
        y_{dd,q}(\tau, \nu) & \hspace{-3mm} = & \hspace{-3mm} \sum\limits_{s=1}^U h_{\mbox{\scriptsize{eff}},q,s}(\tau, \nu) *_{\sigma} x_{dd,s}(\tau, \nu) \, + \, n_{dd,q}(\tau, \nu) \nonumber \\
       h_{\mbox{\scriptsize{eff}},q,s}(\tau, \nu) &  \Define & w_{rx,q}(\tau, \nu) \, *_{\sigma}  h_s(\tau, \nu)  *_{\sigma} \, w_{tx,s}(\tau, \nu).
    \end{eqnarray}Using the expression for $x_{dd,s}(\tau, \nu)$ from (\ref{eqn5}) we get
    \begin{eqnarray}
\label{eqn2113}
   y_{dd,q}(\tau, \nu) & \hspace{-3mm} = & \hspace{-3mm} \sum\limits_{s=1}^U \sum\limits_{k=0}^{M_s -1 }\sum\limits_{l=0}^{N_s -1} x_s[k,l] \, {\Tilde h}_{q,s}^{k,l}(\tau, \nu)  \, + \, n_{dd,q}(\tau, \nu), \nonumber \\
   {\Tilde h}_{q,s}^{k,l}(\tau, \nu) & \Define & h_{\mbox{\scriptsize{eff}},q,s}(\tau, \nu) *_{\sigma} \phi_{dd,s}^{k,l}(\tau, \nu)
 \end{eqnarray}The continuous DD domain match-filtered signal for the $q$-th UT, (i.e., $y_{dd,q}(\tau, \nu)$) is then sampled on the information lattice $\Lambda_q$ of the $q$-th UT, resulting in the discrete DD domain signal

    {\vspace{-2.5mm}
    \small
    \begin{eqnarray}
        \label{rx_dis_sig_ut}
        y_{dd,q}[k',l'] & \hspace{-3.5mm} = &  \hspace{-3.5mm} y_{dd,q}\left(\tau = \frac{k' \tau_{p,q}}{M_q}, \nu = \frac{l' \nu_{p,q}}{N_q} \right) \nonumber \\
        & \hspace{-3mm} = & \hspace{-3mm} \sum_{s=1}^{U} \sum_{k=0}^{M-1} \sum_{l=0}^{N-1} 
        \Tilde{h}_{q,s}^{k,l}[k',l'] \, x_{s}[k,l]  \,  \,  + n_{dd,q}[k',l'], \nonumber \\
        \Tilde{h}_{ q,s}^{k,l}[k',l'] & \Define & {\Tilde h}_{q,s}^{k,l}\left(\tau = \frac{k' \tau_{p,q}}{M_q}, \nu = \frac{l' \nu_{p,q}}{N_q} \right), \nonumber \\
        n_{dd,q}[k',l'] & \hspace{-2mm} \Define &  \hspace{-2mm} n_{dd,q}\left(\tau = \frac{k' \tau_{p,q}}{M_q}, \nu = \frac{l' \nu_{p,q}}{N_q} \right), k',l' \in {\mathbb Z}.
    \end{eqnarray}
    \normalsize}
    The signal processing at the UTs and at the BS has been illustrated through Fig.~\ref{fig:system_model}.

    Separating the signal received from the $q$-th UT in the R.H.S. of (\ref{rx_dis_sig_ut}) we get the useful term (for detecting information transmitted from $q$-th UT) and the interference from all other UTs (multiuser interference (MUI))
    \begin{eqnarray}
        y_{dd,q}[k',l'] & \hspace{-3.5mm} = &  \hspace{-3.5mm} \underbrace{\sum_{k=0}^{M_q-1} \sum_{l=0}^{N_q-1} \Tilde{h}_{q,q}^{k,l}[k',l'] \,\, x_q[k,l]}_{\mbox{useful signal for q-th UT}}  \nonumber \\
        & & \hspace{-20mm} + \, \underbrace{\hspace{-2mm} \sum_{s=1, s \ne q}^{U} \sum_{k=0}^{M_s-1} \sum_{l=0}^{N_s-1} \Tilde{h}_{q,s}^{k,l}[k',l'] \,\, x_s[k,l]}_{\mbox{MUI for q-th UT}} + n_{dd,q}[k',l'].
    \end{eqnarray}Further, for the useful signal term of the $q$-th UT, from the I/O relation derivation in prior works on single-user Zak-OTFS \cite{otfsbook}, we know that the useful signal can be expressed as the discrete twisted convolution between an effective discrete DD channel filter $h_{\mbox{\scriptsize{eff}},q,q}[k',l']$ and a discrete DD domain quasi-periodic information signal $x_{dd,q}[k',l']$, i.e.
    \begin{eqnarray}
    \label{eqn24245}
        \sum_{k=0}^{M_q-1} \sum_{l=0}^{N_q-1} \Tilde{h}_{q,q}^{k,l}[k',l'] \,\, x_q[k,l] \nonumber \\
        & & \hspace{-46mm} = h_{\mbox{\scriptsize{eff}},q,q}[k',l'] \, *_{\sigma} \,  x_{dd,q}[k',l']  \nonumber \\
        & & \hspace{-46mm}  = \sum\limits_{k,l \in {\mathbb Z}} h_{\mbox{\scriptsize{eff}},q,q}[k,l] \, x_{dd,q}[k' - k,l' - l] \, e^{j \frac{2 \pi}{M_q N_q} l (k' - k)} \nonumber \\
        & & \hspace{-46mm} h_{\mbox{\scriptsize{eff}},q,q}[k',l'] \, \Define  \, h_{\mbox{\scriptsize{eff}},q,q}\left(\tau = \frac{k' \tau_{p,q}}{M_q}, \nu = \frac{l' \nu_{p,q}}{N_q} \right) \nonumber \\
        & & \hspace{-46mm}  x_{dd,q}[k',l']  = e^{j 2 \pi \frac{\left\lfloor \frac{k'}{M_q}\right\rfloor l'}{N_q}} \,  x_q[k' \, \mbox{\scriptsize{mod}} \,  M_q, l' \, \mbox{\scriptsize{mod}} \,  N_q],
    \end{eqnarray}where for any two integers $\alpha, \beta$, $\alpha \, \mbox{\scriptsize{mod}} \, \beta$ is the smallest non-negative integer such that $(\alpha - \alpha \, \mbox{\scriptsize{mod}} \,  \beta)$ is an integer multiple of $\beta$. The quasi-periodic information signal $x_{dd,q}[k',l']$ can be expressed in terms of $x_q[k,l]$ also as

    {\vspace{-4mm}
    \small
    \begin{eqnarray}
    \label{eqn25256}
        x_{dd,q}[k',l'] & \hspace{-4mm} = & \hspace{-4mm} \sum\limits_{k=0}^{M_q-1}\sum\limits_{n=0}^{N_q-1}  \sum\limits_{n,m \in {\mathbb Z}} \hspace{-2mm} {\Bigg (} x_q[k,l] \, \delta[k' - k - nM_q]  \nonumber \\
        & & \hspace{18mm} \delta[l' - l - mN_q] \, e^{j 2 \pi n \frac{l}{N_q}} {\Bigg )}.
    \end{eqnarray}\normalsize}

	\section{Proposed Zak-OTFS Multiuser Uplink (Zak-OTFS-MUL) }\label{secproposedMA}
\begin{figure}
     \centering
          \includegraphics[scale=0.27]{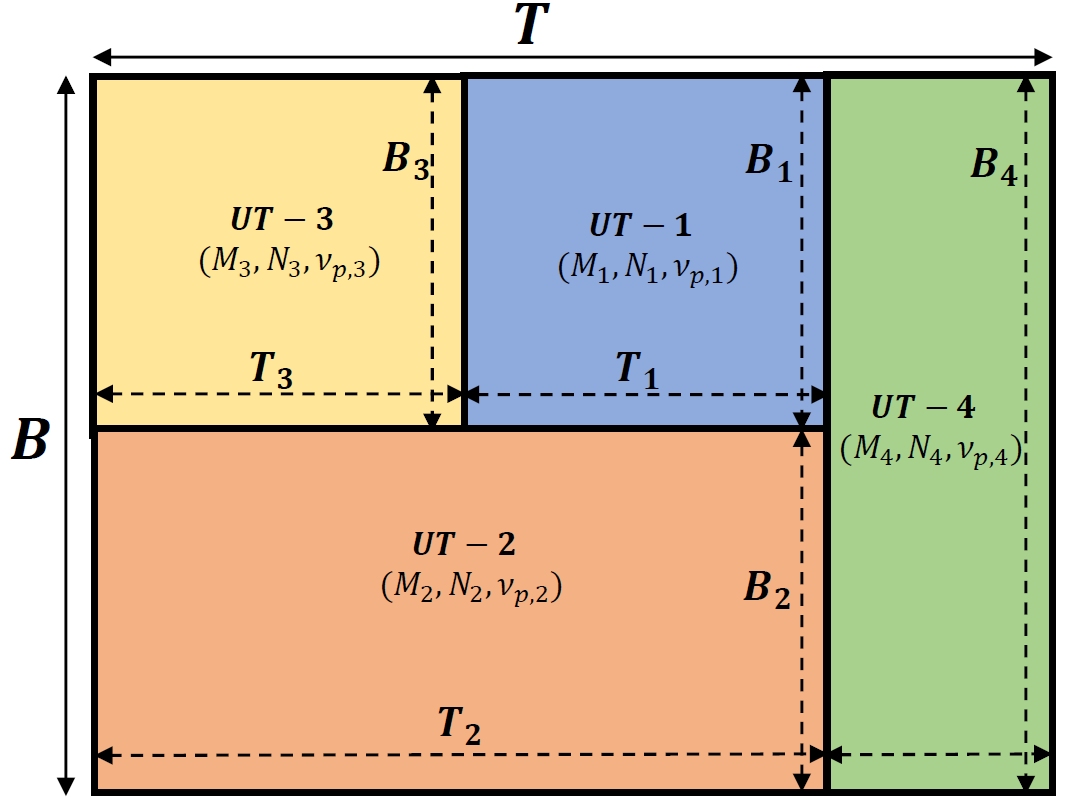}
			\caption{Non-overlapping TF resource allocation for UTs.}
			\label{fig:MA_example}
    \end{figure}

    In this paper, we propose a Zak-OTFS based multiuser uplink system with $U$ UTs. The multiuser system operates within a total bandwidth of $B$ Hz and a time duration of $T$ seconds. Each user is allocated non-overlapping TF resources to achieve orthogonal MA. The allocated time duration $T_s$ and bandwidth $B_s$ for $s$-th UT may vary depending upon its data rate requirements and other usage considerations. Fig. \ref{fig:MA_example} illustrates an example of four UTs assigned distinct non-overlapping TF resources. The DD period ($\tau_{p,s}, \nu_{p,s}$) for the $s$-th UT is chosen to ensure that the crystallization condition is satisfied for the $s$-th UT. The crystallization condition holds for the $s$-th UT if \cite{ZAKOTFS1, ZAKOTFS2, otfsbook}

    \begin{eqnarray}
    \label{cryst_condition}
        \tau_{p,s} > \tau_{max, s} \hspace{0.25cm} \text{and} \hspace{0.25cm} \nu_{p,s} > 2\nu_{max, s}, 
    \end{eqnarray}
    where $\tau_{max, s}$ and $\nu_{max, s}$ are the maximum delay and Doppler spread of the channel for the $s$-th UT, respectively. This condition ensures that the I/O relation for the $s$-th UT is predictable and non-fading \cite{ZAKOTFS2, otfsbook}. Predictability means that the channel response to any pulsone carrier $\psi_s^{k,l}(t)$, $k=0,1,\cdots, M_s -1$, $l=0,1,\cdots, N_s -1$ can be accurately predicted/estimated from the channel response to a particular carrier (e.g., a pilot carrier $\psi_s^{k_p,l_p}(t)$ corresponding to a DD pulse located at $(k_p, k_p)$).
    This ability of Zak-OTFS-MUL to allow each user to choose its own DD period $(\tau_{p,s}, \nu_{p,s})$ (independent of other users) to satisfy the crystallization condition for its own channel make it \emph{more flexible} than multiuser uplink OFDMA where all users need to transmit carriers with the same sub-carrier spacing. 
    
    For the signals of different UTs to occupy non-overlapping regions in the TF domain, in the next section we propose a novel DD domain filtering based method.

    \subsection{Proposed DD domain processing at transmitter}\label{subsecDDprocessing}
    In this section, we propose novel DD domain filtering which ensures that the transmitted signal from each UT is precisely confined within its designated TF allocation.
From (\ref{eqn132}) and (\ref{eqn15}) we know that  the time duration and bandwidth of the Zak-OTFS modulated transmit signal for the $s$-th UT are respectively determined by the time duration of $W_{T_s}(t) = \int w_{T_s}(\nu) e^{j 2 \pi \nu t} \, d\nu$ and the bandwidth of $w_{B_s}(t)$, where $w_{tx,s}(\tau, \nu) = w_{T_s}(\nu) \, w_{B_s}(\tau)$ is the transmit pulse shaping filter. Since any signal cannot be strictly limited in both time and bandwidth, we consider pulse shaping fillters $w_{tx,s}(\tau, \nu)$ where $W_{T_s}(t)$ is approximately limited to the time interval $\left[ - \frac{T_s}{2} \,,\, \frac{T_s}{2} \right]$ and
$W_{B_s}(f) = \int w_{B_s}(\tau) \, e^{-j 2 \pi f \tau} \, d\tau$
is approximately limited to the frequency domain interval $\left[ - \frac{B_s}{2} \,,\, \frac{B_s}{2} \right]$. In general we consider pulse shaping filters which have unit energy, i.e.,
\begin{eqnarray}
\int \vert w_{T_s} (\nu) \vert^2 \, d\nu & = & \int \vert w_{B_s}(\tau) \vert^2 d\tau \, = \, 1,
\end{eqnarray}and are approximately time and bandwidth limited to $T_s$ seconds and $B_s$ Hz respectively, i.e.

{\vspace{-4mm}
\small
\begin{eqnarray}
\label{eqncnstr}
\int\limits_{- \frac{T_s}{2}}^{\frac{T_s}{2}} \vert W_{T_s}(t) \vert^2 \, dt & = & (1 - \epsilon_t) \,,\, 
\int\limits_{- \frac{B_s}{2}}^{\frac{B_s}{2}} \vert W_{B_s}(f) \vert^2 \, df  =  (1 - \epsilon_f) , \nonumber \\
\end{eqnarray}\normalsize}where $\epsilon_t, \epsilon_f > 0$ and are much smaller than one. An example of such a pulse shaping filter is the sinc filter

{\vspace{-4mm}
\small
    \begin{eqnarray}
    \label{sinc}
        {w}_{tx,s}(\tau, \nu) = \underbrace{\sqrt{B_s}\,\text{sinc}(B_s\tau)}_{w_{B_s}(\tau)} \times \underbrace{\sqrt{T_s} \, \text{sinc}(T_s\nu)}_{w_{T_s}(\nu)},
    \end{eqnarray}\normalsize}since for this filter

    {\vspace{-4mm}
    \small
    \begin{eqnarray}
    \label{sinc_FT}
        W_{T_s}(t) & \hspace{-2.5mm} =& \hspace{-2.5mm} \frac{1}{\sqrt{T_s}} \text{rect}\left(\frac{t}{T_s}\right) \,,\,
        W_{B_s}(f)  =  \frac{1}{\sqrt{B_s}} \text{rect}\left(\frac{f}{B_s}\right),
    \end{eqnarray}\normalsize}where the function $\text{rect}(x), x \in {\mathbb R}$ is given by

    {\vspace{-4mm}
    \small
    \begin{eqnarray}
    \text{rect}(x) & \Define 
\begin{cases}
1 &, -\frac{1}{2} \leq x < \frac{1}{2} \\
0 &, \mbox{\scriptsize{otherwise}} \\
\end{cases}.
   \end{eqnarray}\normalsize}
        \begin{figure*}
    \centering
          \includegraphics[scale=0.37]{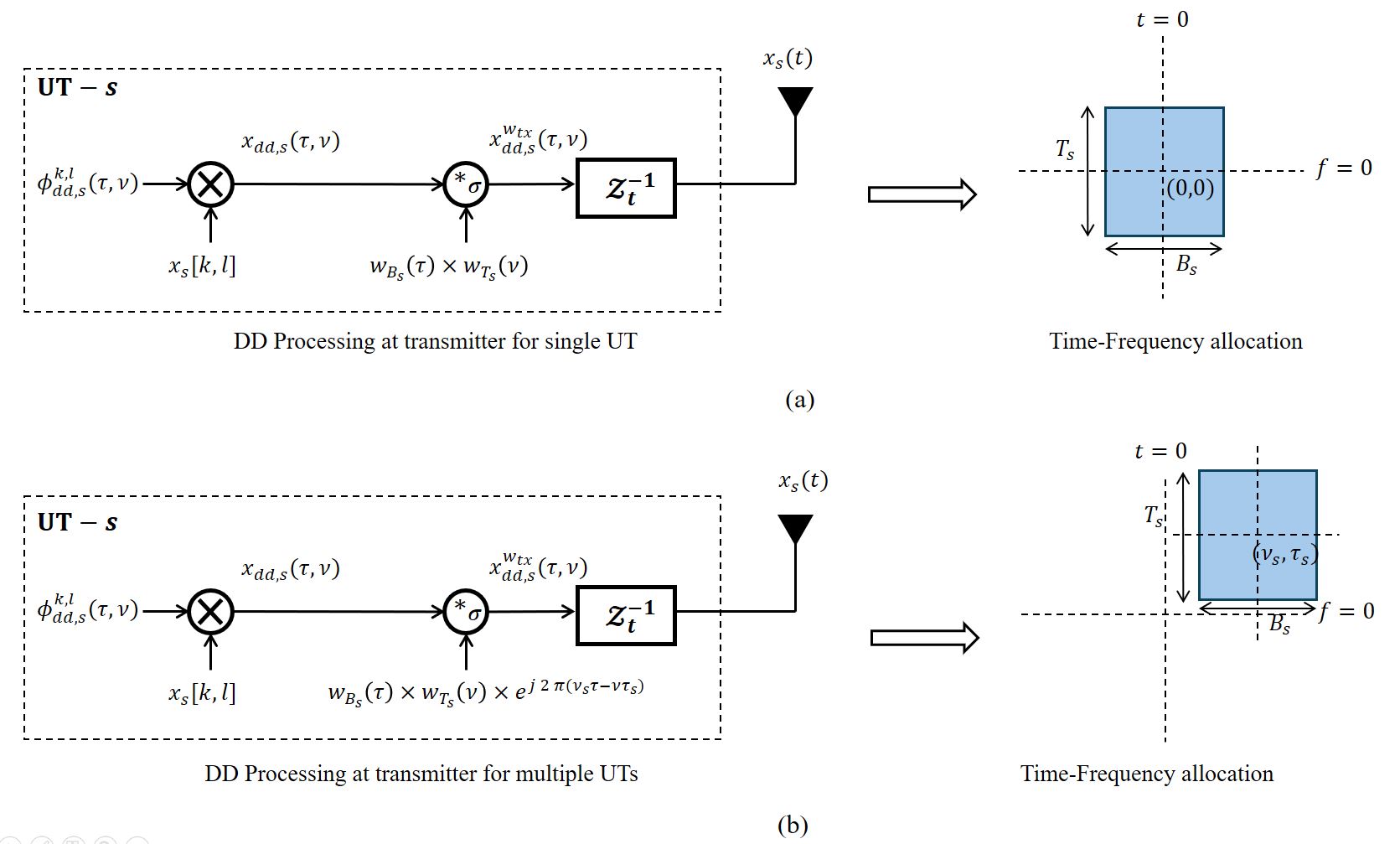}
			\caption{DD domain processing  and corresponding TF allocation : (a) Single UT (b) Multiple UTs}
			\label{fig:DDprocessing}
    \end{figure*}
Substituting the expression for the sinc pulse shaping filter from (\ref{sinc}) in (\ref{eqn14}), the carrier waveform for the $(k,l)$-th information symbol is given by

    {\small
        \begin{eqnarray}
        \label{sinc_psi_t}
            {\psi}^{k,l}_{s}(t) & \hspace{-3.5mm} = &  \hspace{-3.5mm} \sqrt{\frac{\tau_p B_s}{T_s}} \hspace{-1mm}  \sum_{n=-\infty}^{\infty} \hspace{-2mm} \text{sinc}\left [ \hspace{-0.5mm} B_s \hspace{-1mm} \left( \hspace{-0.5mm} t - n\tau_{p,s} - \frac{k \tau_{p,s}}{M_s} \hspace{-0.5mm} \right) \hspace{-0.7mm} \right ] \hspace{-1mm} e^{j 2 \pi \frac{nl}{N_s}} \nonumber \\ 
            &&  \hspace{25mm} \times \hspace{1mm}  \text{rect}\left(\hspace{-1mm} \frac{n\tau_{p,s} + \frac{k\tau_{p,s}}{M_s}}{T_s}\hspace{-1mm}\right),
        \end{eqnarray}
        \normalsize}and its frequency domain realization (Fourier transform) is given by

    {\small
        \begin{eqnarray}
         \label{sinc_psi_f}
            {\Psi}^{k,l}_{s}(f) & \hspace{-3.5mm} = &  \hspace{-3.5mm} \sqrt{\frac{\nu_p T_s}{B_s}} \hspace{-1mm} \left[ \hspace{-1mm} \sum_{m=-\infty}^{\infty} \hspace{-3mm} \text{sinc} \hspace{-1.0mm} \left [ \hspace{-0.5mm} T_s \hspace{-1mm} \left( \hspace{-0.5mm} f \hspace{-0.5mm} - \hspace{-0.5mm} m\nu_{p,s} \hspace{-0.8mm} - \hspace{-0.8mm}\frac{l \nu_{p,s}}{N_s} \hspace{-0.5mm} \right) \hspace{-0.7mm} \right ] \hspace{-1mm} e^{-j 2 \pi ( \frac{mkN_s+lk}{M_sN_s})} \hspace{-1mm} \right] \nonumber \\ 
            && \hspace{48mm} \times \hspace{1mm} \text{rect}\left( \hspace{-1mm}\frac{f}{B_s}\hspace{-1mm}\right).\,\,\,\,\,\hspace{-2.5mm}
        \end{eqnarray}
        \normalsize}
    From (\ref{sinc_psi_t}) and (\ref{sinc_psi_f}), it is evident that $\text{rect}(.)$ approximately limits the signal $\psi^{k,l}_{s}(t)$ to {\small $\left[ -\tfrac{T_s}{2}, \tfrac{T_s}{2}\right)$} seconds in time and {\small $\left[ -\tfrac{B_s}{2}, \tfrac{B_s}{2}\right)$} Hz in bandwidth.\footnote{\footnotesize{From (\ref{sinc_psi_f}) it is clear that the $(k,l)$-th carrier waveform $\psi_s^{k,l}(t)$ is exactly band-limited to $[-\frac{B_s}{2} \,,\, \frac{B_s}{2} ]$, and from (\ref{sinc_psi_t}) it follows that a very small fraction of energy leaks outside the time interval $\left[  -\frac{T_s}{2} \,,\, \frac{T_s}{2} \right]$. The time spread of this leakage beyond the interval $\left[  -\frac{T_s}{2} \,,\, \frac{T_s}{2} \right]$ is $O\left(\frac{1}{B_s}\right)$. }}
In general, for any pulse shaping filter satisfying (\ref{eqncnstr}), the time-spread and frequency-spread of the energy leakage beyond the allocated TF interval $\left[  -\frac{T_s}{2} \,,\, \frac{T_s}{2} \right] \, \times \, \left[  -\frac{B_s}{2} \,,\, \frac{B_s}{2} \right]$ is $O\left( \frac{1}{B_s}\right)$ and $O\left( \frac{1}{T_s}\right)$ respectively. This is clear from (\ref{eqn15}) and (\ref{eqn187635}) given that the spreads of the factors $w_{B_s}(\tau)$ and $w_{T_s}(\nu)$ are $O\left( \frac{1}{B_s}\right)$ and $O\left( \frac{1}{T_s}\right)$ respectively.

    However, with the aforementioned transmit pulse shaping filter,
the transmit signals of all UTs would be located around $t=0$ and $f=0$ respectively in time and frequency domains, and would therefore interfere strongly with each other, as shown in Fig.~\ref{fig:DDprocessing}(a).

Therefore, in this paper we propose a novel DD domain pulse shaping filter design which shifts/translates the TF occupancy of the signal transmitted by a UT by any arbitrary time and frequency shift.

    \begin{theorem}
        \label{theorem2}
        Let the transmitted signal ${x}_{s}(t)$, corresponding to a transmit pulse ${w}_{tx,s}(\tau, \nu) = w_{T_s}(\nu) \, w_{B_s}(\tau)$, occupy the TF region 
        \begin{eqnarray}
            \left[-\frac{T_s}{2}, \frac{T_s}{2}\right) \times \left[-\frac{B_s}{2}, \frac{B_s}{2}\right). \nonumber 
        \end{eqnarray}Then, for the proposed transmit pulse 
        \begin{eqnarray}
        \label{tx_filter_mod}
            {\Tilde w}_{tx,s}(\tau, \nu) = {w}_{tx,s}(\tau, \nu)e^{j2\pi({\nu}_{s} \tau - \nu {\tau}_{s})} 
        \end{eqnarray} 
        the corresponding transmitted  signal ${\Tilde x}_{s}(t)$ occupies the shifted TF region 
        \begin{eqnarray}
            {\tiny \left[-\frac{T_s}{2} + \tau_s, \frac{T_s}{2} + \tau_s \right) \times \left[-\frac{B_s}{2} + \nu_s, \frac{B_s}{2} + \nu_s \right)} \nonumber
        \end{eqnarray} in the TF domain, i.e., a time shift by $\tau_s$ seconds and a frequency shift by $\nu_s$ Hz.
    \end{theorem}

    \begin{IEEEproof}
    See Appendix \ref{appendix_theorem_2}.
    \end{IEEEproof}

    The transmitter signal processing for the proposed pulse shaping filter ($w_{B_s}(\tau) \, w_{T_s}(\nu) \, e^{j 2 \pi (\nu_s \tau - \nu \tau_s)}$ for the $s$-th UT) which confines a UT's signal to its allocated TF region is illustrated in Fig.~\ref{fig:DDprocessing}(b) (shown for the $s$-th UT). Since the TF shift $(\tau_s, \nu_s)$ can be chosen freely, the proposed Zak-OTFS-MUL system allows for \emph{flexible TF resource allocation} and hence it can achieve any TF allocation that is possible in 3GPP 5G NR.
Appropriate selection of the time and frequency shifts $\tau_s, \nu_s, s=1,2,\cdots, U$ ensures that the $U$ UTs have non-overlapping TF allocation within the total system time and bandwidth resource (as for example shown in Fig. \ref{fig:MA_example} for $U=4$ UTs, where the total system time and bandwidth resource is $T$ seconds and $B$ Hz).

    \subsection{DD domain receiver processing at the BS} \label{DDprocessingatRx}
    In this section, we present the DD domain processing required at the BS to extract the discrete DD domain signal for the $q$-th UT. For the proposed filtering at the transmitter, we derive the expression for the effective channel filter $h_{\mbox{\scriptsize{eff}},q,s}(\tau, \nu)$ between the DD signal $x_{dd,s}(\tau, \nu)$ transmitted by the $s$-th UT and the match-filtered signal $y_{dd,q}(\tau, \nu)$ at the BS receiver.
    This then gives the expression for the discrete DD domain I/O relation between the sampled match-filtered signal for the $q$-th UT i.e., $y_{dd,q}[k,l]$ and the information symbols $x_s[k,l]$ transmitted by the $s$-th UT.

    At the BS, information symbols transmitted by the $q$-th UT are detected from the sampled matched-filter output (see (\ref{rxmatchfilt}) and (\ref{rx_dis_sig_ut})).
    For a single-user Zak-OTFS based system, the receive filter which
    optimizes the SNR at the match filter output is  
    $w_{rx, q}(\tau, \nu)  =    w^{*}_{tx,q}(-\tau, -\nu) e^{j 2 \pi \nu \tau}$ \cite{Hanly23,ZAKOTFS1, ZAKOTFS2, otfsbook}.
    
    For the proposed transmit pulse shaping filter $w_{tx,q}(\tau, \nu) = w_{B_q}(\tau) \, w_{T_q}(\nu) \, e^{j 2 \pi (\nu_q \tau - \nu \tau_q)}$ (see Theorem \ref{theorem2}), and therefore the corresponding receive match-filter is given by
    \begin{eqnarray}
    \label{match_filter}
       w_{rx, q}(\tau, \nu) & \hspace{-2.5mm} = &  \hspace{-2.5mm} w^{*}_{B_q}(-\tau) w^{*}_{T_q}(-\nu) e^{j2\pi(\nu_q\tau-\nu\tau_q)} e^{j 2 \pi \nu\tau}.
    \end{eqnarray}
    Substituting (\ref{match_filter}) into (\ref{eqn2020}), the effective continuous DD domain channel filter for the channel between the match-filtered signal $y_{dd,q}(\tau, \nu)$ for the $q$-th UT and the transmitted DD signal $x_{dd,s}(\tau, \nu)$ from the $s$-th UT is given by

    {\small
    \begin{eqnarray}
        \label{effcontch}
        h_{\mbox{\scriptsize{eff}},q,s}(\tau, \nu) & \hspace{-3.5mm} = &  \hspace{-3.5mm} {\Big (} w^{*}_{B_q}(-\tau) w^{*}_{T_q}(-\nu) e^{j2\pi(\nu_q\tau-\nu\tau_q)} e^{j 2 \pi \nu\tau} {\Big )} \nonumber \\
        & & \hspace{-10mm} *_{\sigma} \, h_{s}(\tau, \nu) \, *_{\sigma} \, {\Big (}  w_{B_s}(\tau) \, w_{T_s}(\nu) \, e^{j 2 \pi (\nu_s \tau - \nu \tau_s)}{\Big )}.
    \end{eqnarray}
    \normalsize}We assume that the wireless channel $h_{s}(\tau, \nu)$ between the $s$-th UT and the BS consists of $P_s$ propagation paths, where the delay and Doppler shift associated with the $i$-th path are $\tau_{i,s}$ and $\nu_{i,s}$ respectively. The maximum delay and Doppler spreads are denoted by $\tau_{max,s}$ and $\nu_{max,s}$ respectively. Then
    {
    \begin{eqnarray}
        \label{wirelessch}
        h_{s}(\tau, \nu) & \hspace{-1.5mm} = &  \hspace{-1.5mm}  \sum_{i=1}^{P_s} h_{i,s} \delta(\tau - {\tau}_{i,s}) \delta(\nu - {\nu}_{i,s}).
    \end{eqnarray}
    \normalsize}
    The expression of the effective DD domain continuous channel $h_{\mbox{\scriptsize{eff}},q,s}(\tau, \nu)$ is given by the following theorem.
    \begin{theorem}
        \label{theorem4}
        The effective continuous DD channel $h_{\mbox{\scriptsize{eff}},q,s}(\tau, \nu)$ in (\ref{effcontch}) is given by
    {
    \begin{eqnarray}
        \label{effcontchexp}
       h_{\mbox{\scriptsize{eff}},q,s} (\tau, \nu)& \hspace{-3.5mm} = &  \hspace{-3.5mm}  \sum_{i=1}^{P_s} h_{i,s} e^{j 2 \pi \nu_s(\tau - \tau_{i,s})} e^{j 2 \pi \nu_{i,s}(\tau +\tau_s - \tau_{i,s})} e^{-j2\pi\nu\tau_s} \nonumber \\
        && \hspace{20mm} \times \zeta_{q,s,i}(\tau) \times \eta_{q,s,i}(\tau, \nu) \\
        \text{where} \nonumber
    \end{eqnarray}
    \normalsize} 

    {\small
    \begin{eqnarray}
       \zeta_{q,s,i}(\tau) & \hspace{-3.5mm} = &  \hspace{-3.5mm}    \int w^{*}_{B_q}(-\tau') w_{B_s}(\tau - \tau'- \tau_{i,s}) e^{j 2 \pi (\nu_q- \nu_s - \nu_{i,s})\tau'} d\tau' \nonumber
    \end{eqnarray} 
    \normalsize}

    {\small
    \begin{eqnarray}
       \eta_{q,s,i}(\tau, \nu) & \hspace{-3.5mm} = &  \hspace{-3.5mm}    \int w^{*}_{T_q}(-\nu') w_{T_s}(\nu - \nu'- \nu_{i,s}) e^{j 2 \pi \nu'(\tau - (\tau_q-\tau_s))} d\nu' \nonumber
    \end{eqnarray} 
    \normalsize}

    \end{theorem}
    \begin{IEEEproof}
	See Appendix \ref{appendix_theorem_4}.
	\end{IEEEproof}
Substituting the expression of $h_{\mbox{\scriptsize{eff}},q,s} (\tau, \nu)$ from (\ref{effcontchexp}) into (\ref{eqn2113}) gives the expression for ${\Tilde h}_{q,s}^{k,l}(\tau, \nu)$. Sampling ${\Tilde h}_{q,s}^{k,l}(\tau, \nu)$ on the information grid of the $q$-th UT gives the expression for $\Tilde{h}_{ q,s}^{k,l}[k',l']$ in (\ref{hcoeff}) (see top of this page).
The I/O relation between the match-filtered output $y_{dd,q}[k',l']$ and the information symbols $x_s[k,l]$ of the $s$-th UT is given in (\ref{rx_dis_sig_ut}) with $\Tilde{h}_{ q,s}^{k,l}[k',l']$ given by (\ref{hcoeff}).


    \begin{figure*}
			{\small
            \vspace{-5mm}
				\begin{eqnarray}\label{hcoeff}
        \Tilde{h}_{ q,s}^{k,l}[k',l'] & \hspace{-2.5mm} = &  \hspace{-2.5mm} h_{\mbox{\scriptsize{eff}},q,s} (\tau, \nu) *_{\sigma} \phi^{k,l}_{dd} (\tau, \nu) \big\vert_{\left( \hspace{-1mm} \tau = \frac{k'\tau_{p,q}}{M_q}, \nu = \frac{l'\nu_{p,q}}{N_q} \hspace{-1mm} \right)} \nonumber \\
        \hspace{-3.5mm}& \hspace{-3.5mm} = &  \hspace{-3.5mm}  \sum_{n,m}  h_{\mbox{\scriptsize{eff}},q,s}\hspace{-0.5mm}\left(\hspace{-0.75mm}\frac{k' \tau_{p,q}}{M_q} - n\tau_{p,s} - \frac{k \tau_{p,s}}{M_s}  , \frac{l'\nu_{p,q}}{N_q} - m\nu_{p,s} - \frac{l \nu_{p,s}}{N_s} \hspace{-0.75mm}\right)  \times e^{j 2 \pi \frac{nl}{N_s}}e^{j 2 \pi \left( \frac{l'\nu_{p,q}}{N_q} - m\nu_{p,s} - \frac{l\nu_{p,s}}{N_s}\right)\left(n \tau_{p,s} + \frac{k\tau_{p,s}}{M_s}\right)}.
    \end{eqnarray}
			}
			{\vspace{-4mm}
				\begin{eqnarray*}
					\hline
			\end{eqnarray*}}
				\normalsize
	\end{figure*}

    \subsection{TF localization of a UT's transmit signal}
\label{tflocal}
So far, we have proposed transmit pulse shape filtering for each UT which ensures that the transmit signal is localized to the TF region allocated to that UT. In an ideal scenario (no delay and Doppler spread) we therefore expect to have negligible multiuser interference in the match-filtered signal at the receiver.

In (\ref{rxmatchfilt}) we have seen that to detect the information transmitted by the
$q$-th UT we match-filter the received DD domain signal $y_{dd}(\tau, \nu)$ with the receive filter $w_{rx,q}(\tau, \nu)$ of the $q$-th UT resulting in the match-filtered signal $y_{dd,q}(\tau, \nu)$. From (\ref{eqn2113}) it is clear that this match-filtered signal contains contributions from the signal transmitted by the $q$-th UT as well as those transmitted by other UTs. To be precise, the interference from the $s$-th UT is given by the term $x_s[k,l] \, {\Tilde h}_{q,s}^{k,l}(\tau, \nu)$ in the R.H.S. of (\ref{eqn2113}), where
${\Tilde h}_{q,s}^{k,l}(\tau, \nu) = h_{\mbox{\scriptsize{eff}},q,s}(\tau, \nu) *_{\sigma} \phi_{dd,s}^{k,l}(\tau, \nu)$.

To assess the TF localization of the signal transmitted by the $s$-th UT, we therefore examine the energy of the interference to the received match-filtered signal for the $q$-th UT i.e., $y_{dd,q}(\tau, \nu)$ from the $(k,l)$-th information symbol $x_s[k,l]$ transmitted
by the $s$-th UT. This interference is given by the term $x_s[k,l] {\Tilde h}_{q,s}^{k,l}(\tau, \nu)$ in (\ref{eqn2113}) and its average energy is\footnote{\footnotesize{Note that ${\Tilde h}_{q,s}^{k,l}(\tau, \nu) = h_{\mbox{\scriptsize{eff}},q,s}(\tau, \nu) *_{\sigma} \phi_{dd,s}^{k,l}(\tau, \nu)$ is a quasi-periodic signal since $\phi_{dd,s}^{k,l}(\tau, \nu)$ is quasi-periodic and twisted convolution preserves quasi-periodicity \cite{otfsbook}. Further, for any quasi-periodic signal $b_{dd}(\tau, \nu)$ with delay and Doppler period $\tau_p$ and $\nu_p = 1/\tau_p$ respectively, and having TD realization (i.e., inverse Zak-transform of $b_{dd}(\tau, \nu)$) $b(t) = {\mathcal Z}_t^{-1}\left( b_{dd}(\tau, \nu) \right)$, the signal energy is given by $\int\limits_{-\infty}^{\infty} \vert b(t) \vert^2 \, dt \, = \, \int\limits_{0}^{\tau_p} \int\limits_{0}^{\nu_p} \left\vert b_{dd}(\tau, \nu) \right\vert^2 \, d\tau, d\nu$  \cite{otfsbook}.}}
\begin{eqnarray}
{\mathbb E}\left[\vert x_s[k,l] \vert^2 \right] \int\limits_{0}^{\tau_{p,s}}\int\limits_{0}^{\nu_{p,s}} \left\vert {\Tilde h}_{q,s}^{k,l}(\tau, \nu) \right\vert^2 \, d\tau \, d\nu.
\end{eqnarray}On the other hand the average energy of $x_s[k,l]$ in the match-filtered output for the $s$-th user (i.e., useful signal of $s$-th user) is
\begin{eqnarray}
{\mathbb E}\left[ \vert x_s[k,l] \vert^2 \right] \int\limits_{0}^{\tau_{p,s}}\int\limits_{0}^{\nu_{p,s}} \left\vert {\Tilde h}_{s,s}^{k,l}(\tau, \nu) \right\vert^2 \, d\tau \, d\nu.
\end{eqnarray}
With i.i.d. information symbols $x_s[k,l]$,$k=0,1,\cdots, M_s -1$, $l=0,1,\cdots, N_s -1$, the total received average interference energy at the match-filtered output for the $q$-th UT from the transmission of all $M_s N_s$ symbols of the $s$-th UT is
\begin{eqnarray}
\label{eqnIqs}
I_{q,s} & \hspace{-2.5mm} \Define & \hspace{-2.5mm} {\mathbb E}\left[ \vert x_s[k,l] \vert^2 \right] \hspace{-1mm} \sum\limits_{k=0}^{M_s -1 }   \sum\limits_{l=0}^{N_s -1 }  \hspace{-1mm}  \int\limits_{0}^{\tau_{p,s}}\hspace{-1mm} \int\limits_{0}^{\nu_{p,s}} \hspace{-1mm} \left\vert {\Tilde h}_{q,s}^{k,l}(\tau, \nu) \right\vert^2 d\tau d\nu,
\end{eqnarray}when each symbol transmitted by the $s$-th UT has same average energy. Similarly, the average received total useful signal at the match-filtered output of the $s$-th UT is 
\begin{eqnarray}
S_{s,s} & \hspace{-2.5mm} \Define & \hspace{-2.5mm} {\mathbb E}\left[ \vert x_s[k,l] \vert^2 \right] \hspace{-1mm} \sum\limits_{k=0}^{M_s -1 }   \sum\limits_{l=0}^{N_s -1 }  \hspace{-1mm}  \int\limits_{0}^{\tau_{p,s}}\hspace{-1mm} \int\limits_{0}^{\nu_{p,s}} \hspace{-1mm} \left\vert {\Tilde h}_{s,s}^{k,l}(\tau, \nu) \right\vert^2 d\tau \, d\nu.
\end{eqnarray}We are interested in the ratio
\begin{eqnarray}
\label{eqnratio45}
\frac{I_{q,s}}{S_{s,s}} & = & \frac{\sum\limits_{k=0}^{M_s -1 }   \sum\limits_{l=0}^{N_s -1 }   \int\limits_{0}^{\tau_{p,s}}\int\limits_{0}^{\nu_{p,s}} \left\vert {\Tilde h}_{q,s}^{k,l}(\tau, \nu) \right\vert^2 \, d\tau \, d\nu}{\sum\limits_{k=0}^{M_s -1 }   \sum\limits_{l=0}^{N_s -1 }   \int\limits_{0}^{\tau_{p,s}}\int\limits_{0}^{\nu_{p,s}} \left\vert {\Tilde h}_{s,s}^{k,l}(\tau, \nu) \right\vert^2 \, d\tau \, d\nu}.
\end{eqnarray}In this section, we are interested in approximately quantifying the ratio  $I_{q,s}/S_{s,s}$ when the channel of the $s$-th UT is ideal (i.e., $h_s(\tau,\nu) = \delta(\tau) \, \delta(\nu)$) as it gives us a measure of the TF localization of the signal transmitted by the $s$-th UT. Later in Section \ref{numsec}, through numerical simulations we also study this ratio for doubly-spread channels to quantify the impact of the channel delay and Doppler spread on this ratio. With $h_s(\tau,\nu) = \delta(\tau) \, \delta(\nu)$, from (\ref{effcontchexp}) we get
\begin{eqnarray}
\label{eqn4612}
h_{\mbox{\scriptsize{eff}},q,s}(\tau, \nu)  & = & e^{j 2 \pi(\nu_s \tau - \nu \tau_s)} \, \zeta_{q,s,1}(\tau) \, \eta_{q,s,1}(\tau, \nu), \nonumber \\
\zeta_{q,s,1}(\tau)  & \hspace{-3mm} = & \hspace{-3mm} \int w^*_{B_q}(-\tau') \, w_{B_s}(\tau - \tau') e^{j 2 \pi (\nu_q - \nu_s) \tau'} d\tau' \nonumber \\
\eta_{q,s,1}(\tau, \nu) & \hspace{-3mm} = & \hspace{-3mm} \int w_{T_q}^*(-\nu') w_{T_s}(\nu - \nu') e^{j 2 \pi \nu'(\tau - (\tau_q - \tau_s))} \, d\nu'. \nonumber \\
\end{eqnarray}Expressing the integral for $\zeta_{q,s,1}(\tau)$ in terms of $W_{B_q}(f)$ and $W_{B_s}(f)$ (see (\ref{eqn26})) we get
\begin{eqnarray}
\label{eqn4848}
\zeta_{q,s,1}(\tau) & = & \int W^*_{B_q}(f - (\nu_q - \nu_s)) \, W_{B_s}(f) \, e^{j 2 \pi f \tau} \, df, \nonumber \\
& & \hspace{-13mm} = \int W^*_{B_q}(f - \nu_q) \, W_{B_s}(f- \nu_s) \, e^{j 2 \pi( f - \nu_s) \tau} \, df
\end{eqnarray}where the second step follows from the first step by substituting the integration variable $f$ with $(f - \nu_s)$. From (\ref{eqncnstr}), we know that both $W_{B_q}(f)$ and $W_{B_s}(f)$ have almost all their energy restricted to the intervals $[-B_q/2 \,,\, B_q/2]$ and $[-B_s/2 \,,\, B_s/2]$ and therefore in the integral in (\ref{eqn4848}), the term $W_{B_q}^*(f - \nu_q)$ is supported in the interval $[\nu_q - B_q/2 \,,\, \nu_q + B_q/2]$ which is same as the FD interval occupied by the signal transmitted by the $q$-th UT (see Theorem \ref{theorem2}). Similarly, the term $W_{B_s}(f - \nu_s)$ is supported in the interval $[\nu_s - B_s/2 \,,\, \nu_s + B_s/2]$ which is same as the frequency domain interval occupied by the signal transmitted by the $s$-th UT. If the $q$-th and $s$-th UT occupy non-overlapping frequency domain intervals (as for example, UT-2 and UT-3 in Fig.~\ref{fig:MA_example}) in which case $\zeta_{q,s,1}(\tau)$ will have a very small value for any $\tau$ and therefore $h_{\mbox{\scriptsize{eff}},q,s}(\tau, \nu)$ will be small which means that the interference energy $I_{q,s}$ will be small.

However, it is possible that these two users have overlapping FD intervals (as for example UT-1 and UT-4 in Fig.~\ref{fig:MA_example}) in which case $\zeta_{q,s,1}(\tau)$ can be significant over some interval of values for $\tau$. This is clear from the expression for $\zeta_{q,s,1}(\tau)$ in (\ref{eqn4612}), where $w_{B_q}(\cdot)$ and $w_{B_s}(\cdot)$
are supported over an interval of spread $O(1/B_q)$ and $O(1/B_s)$ respectively and therefore $\zeta_{q,s,1}(\tau)$ can take significant values only for $\tau \approx O(1/B_q) + O(1/B_s)$. In (\ref{eqn4612}), the integral expression for $\eta_{q,s,1}(\tau, \nu)$ can be written in terms of $W_{T_q}(t)$ and $W_{T_s}(t)$ (see (\ref{eqn15})) as
\begin{eqnarray}
\label{eqn48}
\eta_{q,s,1}(\tau, \nu) &  &  \nonumber \\
& & \hspace{-20mm} = \int W_{T_q}^*(t - \tau_q + \tau) \, W_{T_s}(t - \tau_s) \, e^{-j 2 \pi \nu (t - \tau_s)} \, dt.
\end{eqnarray}Since $W_{T_q}(t)$ and $W_{T_s}(t)$ are limited to the time intervals $[-T_q/2 \,,\, T_q/2]$ and $[-T_s/2 \,,\, T_s/2]$ respectively, it follows that $\eta_{q,s,1}(\tau, \nu)$ can take significant values only for $\tau$ satisfying
\begin{eqnarray}
\label{eqn49}
(\tau_q - \tau_s) - \frac{(T_q + T_s)}{2} \, < \, \tau \, < \, (\tau_q - \tau_s) + \frac{(T_q + T_s)}{2}.
\end{eqnarray}Since two UTs cannot be allocated the same TF region, if two UTs have overlapping FD allocation then their TD allocation must be non-overlapping, i.e., the time intervals $[\tau_q - T_q/2 \,,\, \tau_q + T_q/2]$ and $[\tau_s - T_s/2 \,,\, \tau_s + T_s/2]$ must be non-overlapping. Therefore, either $(\tau_s - \tau_q) > (T_q + T_s)/2$, for which the overlap between the support sets of $W_{T_q}(t - \tau_q + \tau)$ and $W_{T-s}(t - \tau_s)$ is an interval of duration at most $\vert \tau \vert$, or else if $(\tau_q - \tau_s) > (T_q + T_s)/2$ then the overlap interval is also of duration at most $\vert \tau \vert$. 
Since $\zeta_{q,s,1}(\tau)$ takes significant values only for $\tau \approx O(1/B_q) + O(1/B_s)$ the support set of the integrand in the R.H.S. of (\ref{eqn48}) is of size $\vert \tau \vert \approx O(1/B_q) + O(1/B_s)$. The unit energy filter $W_{T_q}(t)$
has its energy spread over a much larger interval of size $T_q$, and therefore the value of $\eta_{q,s,1}(\tau, \nu)$ for $\tau \approx O(1/B_q) + O(1/B_s)$ is at most $ \left(O(1/B_q) + O(1/B_s) \right)/\min(T_q, T_s)$ which is $\approx \left(O(1/(M_q N_q)) + O(1/(M_s N_s)) \right)$ since $B_s T_s = M_s N_s$ and $B_q T_q = M_q N_q$. From the expression of $\eta_{q,s,1}(\tau, \nu)$ in (\ref{eqn4612})
it is also clear that $\eta_{q,s,1}(\tau, \nu)$ takes significant values only for $\vert \nu \vert \approx O(1/T_q) + O(1/T_s)$ since the spread of filters $w_{T_q}(\nu)$ and $w_{T_s}(\nu)$ are $\approx O(1/T_q)$ and $\approx O(1/T_s)$ respectively.

The maximum possible magnitude of $\zeta_{q,s,1}(\tau)$ is one since $w_{B_q}(\tau)$ have unit energy.
This then implies that $\vert h_{\mbox{\scriptsize{eff}},q,s}(\tau, \nu) \vert$ takes values at most $\approx \left(O(1/(M_q N_q)) + O(1/(M_s N_s)) \right)$ only for $\vert \tau \vert \approx O(1/B_q) + O(1/B_s)$, $\vert \nu \vert \approx O(1/T_q) + O(1/T_s)$ and even smaller values elsewhere. Since ${\Tilde h}_{q,s}^{k,l}(\tau, \nu) = h_{\mbox{\scriptsize{eff}},q,s}(\tau, \nu) *_{\sigma} \phi_{dd,s}^{k,l}(\tau, \nu)$ and $\phi_{dd,s}^{k,l}(\tau, \nu)$ (see (\ref{basis_sig})) is just a quasi-periodic Dirac-delta DD pulse at $(k/B_s, l/T_s)$, the DD domain energy distribution of
${\Tilde h}_{q,s}^{k,l}(\tau, \nu)$ (within one DD period) is simply a shifted version of the energy distribution of $h_{\mbox{\scriptsize{eff}},q,s}(\tau, \nu)$. Therefore, for an ideal channel between the BS and the $s$-th UT, the contribution of an information symbol of the $s$-th UT to the interference energy $I_{q,s}$ in the match-filter output for the $q$-th UT is at most ${\mathbb E}\left[ \vert x_s[k,l] \vert^2 \right] \, O(1/K^3)$ where $K = \min(M_s N_s, M_q N_q)$ (since the integrand in (\ref{eqnIqs}) has value $O(1/K^2)$ over an area at most $O(1/K)$ of the fundamental DD period $\{ 0 \leq \tau < \tau_{p,s} \,,\, 0 \leq \nu < \nu_{p,s} \}$). At the same time, since the pulse shaping and match filters have unit energy the contribution of an information symbol $x_s[k,l]$ to the energy of the
useful signal power $S_{s,s}$ at the output of the match-filter for the $s$-th UT is unity. Hence, the ratio $I_{q,s}/S_{s,s}$ in (\ref{eqnratio45}) is at most $\approx O(1/K^3)$ where $K = \min(M_s N_s, M_q N_q)$.

For communication scenarios with moderate/large frames where the amount of physical resource $M_s N_s$ allocated to the $s$-th UT and for that matter any UT is at least hundred, the ratio $I_{q,s}/I_{s,s}$ is at most $\approx O(10^{-6})$, i.e., the interference energy is more than $60$ dB below the useful signal energy.

 In a non-ideal channel, delay and Doppler spread will result in leakage of a user's signal into the match-filtered output of another user allocated adjacent TF resource.
For example, in Fig.~\ref{fig:MA_example}, delay-spread in the channel between UT-1 and BS results in leakage of UT-1's signal into the resource allocated to UT-4. Similarly, Doppler-spread in UT-1's channel results in leakage of UT-1's signal into the resource allocated to UT-2. In the proposed Zak-OTFS-MUL, we do not provision for any guard TF resources separating the resource allocated to users having adjacent time domain or frequency domain resource allocation. Despite this, the proposed pulse-shaping at the transmitter and matched filtering at the BS is such that the effective interference to useful signal ratio ($I_{q,s}/S_{s,s}$ in (\ref{eqnratio45})) is small. This is confirmed through numerical simulations (see Fig.~\ref{MUI_nu_maxUT1} and Fig.~\ref{MUI_nu_maxUT3} in Section \ref{numsec}).

\subsection{Embedded Pilot and Data Frame for Multiple UTs}
\label{embedpilot}
From the expression of the useful received signal in the match-filtered
output $y_{dd,q}[\cdot, \cdot]$ for the $q$-th UT, it is clear that to detect the information symbols $x_q[k,l]$, the BS receiver needs to estimate the taps of the effective discrete DD channel filter $h_{\mbox{\scriptsize{eff}},q,q}[\cdot, \cdot]$ (see (\ref{eqn24245})).

\begin{figure}
     \centering
     \hspace*{-5mm}
          \includegraphics[scale=0.25]{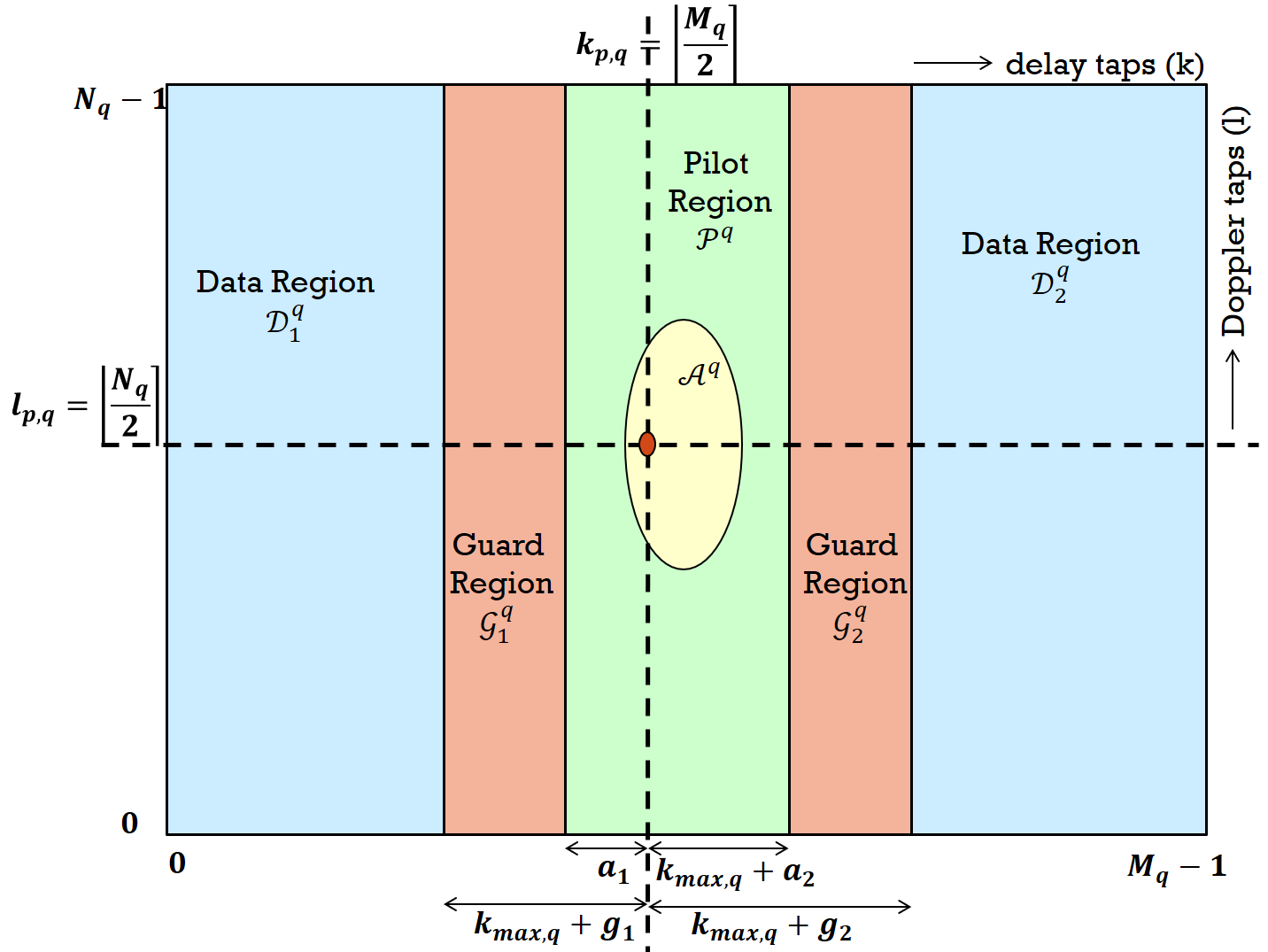}
			\caption{Embedded pilot and data frame for $s$-th UT with data, pilot and guard region}
			\label{fig:modelfree}
    \end{figure}

We adopt the estimation method proposed in \cite{ZAKOTFS2}.
Each UT transmits an embedded pilot and data frame, as illustrated in Fig.~\ref{fig:modelfree}. For the \( q \)-th UT, a pilot is transmitted on a quasi-periodic DD pulse/carrier located at \( (k_{p,q}, l_{p,q}) \), $k_{p,q} \in \{ 0, 1, \cdots, M_q -1 \}, l_{p,q} \in \{ 0, 1, \cdots. N_q -1 \}$ surrounded by the following regions:
\begin{itemize}
    \item the \emph{pilot region} \( \mathcal{P}^q \),
    \item the \emph{guard region} \( \mathcal{G}^q = \mathcal{G}_1^q \cup \mathcal{G}_2^q \), and
    \item the \emph{data region} \( \mathcal{D}^q = \mathcal{D}_1^q \cup \mathcal{D}_2^q \).
\end{itemize}

To minimize interference between data and pilot, the pilot region \( \mathcal{P}^q \) is surrounded by the guard region \( \mathcal{G}^q \). No symbols are transmitted in the region \( \mathcal{P}^q \cup \mathcal{G}^q \), except for a single pilot at location \( (k_{p,q}, l_{p,q}) \). The region denoted by \( \mathcal{A}^q \) represents the support/span of the received pilot signal. The transmit pilot signal is embedded in the discrete DD transmit signal $x_{dd,q}[\cdot, \cdot]$ (see (\ref{eqn25256})) and is given by

{\vspace{-4mm}
\small
\begin{eqnarray}
    x_{dd, p, q}[k',l'] & = & \hspace{-3mm} \sum\limits_{n,m \in {\mathbb Z}} {\Bigg (} \delta[k' - k_{p,q} - nM_q ] \, \delta[l' - l_{p,q} - nN_q ] \nonumber \\
    & & \hspace{10mm} \, e^{j 2 \pi n \frac{l_{p,q}}{N_q} } {\Bigg )}.
\end{eqnarray}\normalsize}The received pilot signal is therefore given by

{\vspace{-4mm}
\small
\begin{eqnarray}
\label{eqn52523}
y_{dd, p, q}[k',l'] & = & h_{\mbox{\scriptsize{eff}},q,q}[k',l'] \, *_{\sigma} \, x_{dd, p, q}[k',l']  \nonumber \\
& & \hspace{-25mm} = \underbrace{h_{\mbox{\scriptsize{eff}},q,q}[k' - k_{p,q},l' - l_{p,q}] \, e^{j 2\pi k_{p,q} \frac{(l' - l_{p,q})}{M_q N_q}}}_{\mbox{\small{has support set}} \, \mathcal{A}^q \, \mbox{\small{in Fig. \ref{fig:modelfree}}} } \nonumber \\
& & \hspace{-25mm} \, + \, \hspace{-6mm} \sum\limits_{\substack{n,m \in {\mathbb Z}, \\ (n,m) \ne (0,0)}} \hspace{-4mm} {\Bigg (} h_{\mbox{\scriptsize{eff}},q,q}[k' - k_{p,q} -nM_q,l' - l_{p,q} - mN_q] \, \nonumber \\
& & \hspace{-10mm} e^{j 2\pi (k_{p,q} + nM_q) \frac{(l' - l_{p,q} - mN_q)}{M_q N_q}} {\Bigg )}.
\end{eqnarray}\normalsize}The first term is simply the taps of the effective discrete DD channel filter $h_{\mbox{\scriptsize{eff}},q,q}[k',l']$ shifted by $(k_{p,q}, l_{p,q})$ and scaled by a deterministic unit modulus scalar. The support set of this term is shown in Fig.~\ref{fig:modelfree}
as ${\mathcal A}^q$. The other terms inside the summation in the RHS of (\ref{eqn52523}) corresponding to $(n,m) \ne (0,0)$ have support sets which are also shifted versions of the support set of $h_{\mbox{\scriptsize{eff}},q,q}[k',l']$ and which do not overlap/alias with ${\mathcal A}_q$ (i.e., no DD domain aliasing) when the crystallization condition is satisfied, i.e., the spread of $h_{\mbox{\scriptsize{eff}},q,q}[k',l']$ along the delay and Doppler axis is less than $M_q$ and $N_q$ respectively. 

From (\ref{eqn24245}) we know that $h_{\mbox{\scriptsize{eff}},q,q}[k',l']$ is simply the effective continuous channel filter $h_{\mbox{\scriptsize{eff}},q,q}(\tau, \nu)$ sampled on the information grid/lattice of the $q$-th UT. From (\ref{effcontch}), $h_{\mbox{\scriptsize{eff}},q,q}(\tau, \nu) = {\Big (} w^{*}_{B_q}(-\tau) w^{*}_{T_q}(-\nu) e^{j2\pi(\nu_q\tau-\nu\tau_q)} e^{j 2 \pi \nu\tau} {\Big )} \,  *_{\sigma} \, h_{q}(\tau, \nu) \, *_{\sigma} \, {\Big (}  w_{B_q}(\tau) \, w_{T_q}(\nu) \, e^{j 2 \pi (\nu_q \tau - \nu \tau_q)}{\Big )}$ and therefore the delay/Doppler domain spread of \( \mathcal{A}^q \) depends on the delay/Doppler spreads of the transmit pulse, the receive match filter, and the maximum delay spread \( \tau_{\text{max},q} \)/maximum Doppler spread \( \nu_{\text{max},q} \) of the channel spreading function \( h_q(\tau, \nu) \). The pilot region \( \mathcal{P}^q \) is designed to fully encompass \( \mathcal{A}^q \).

In our simulations, we consider scenarios with large Doppler spread and hence, the pilot region is shaped as a strip along the Doppler (vertical) domain. In the delay domain, the pilot region spans from \( k_{p,q} - a_1 \) to \( k_{p,q} + k_{\text{max},q} + a_2 \), and the guard region ${\mathcal G}_1^q$ spans from \( k_{p,q} - k_{\text{max},q} - g_1 \) to \( k_{p,q} - a_1 - 1 \), where $k_{\text{max}, q} = \left\lceil B_q \, \tau_{\text{max},q} \right\rceil$. The guard region ${\mathcal G}_2^q$ spans from \( k_{p,q} + k_{\text{max},q} + a_2 + 1 \) to \( k_{p,q} + k_{\text{max},q} + g_2\).
The additional delay bins beyond \( k_{\text{max}, q} \) are included to accommodate the spread of the transmit and receiver filters \( w_{\text{tx}, q}(\tau, \nu) \) and $w_{rx,q}(\tau, \nu)$. For the same reason $a_1$ delay bins are allocated from $(k_{p,q} - a_1)$ to $(k_{p,q} - 1)$ even though the physical channel paths in $h_q(\tau, \nu)$ have non-negative delays.

For the \( q \)-th UT, the transmitted symbols \( x_q[k,l] \) on the embedded pilot-data frame are given by
\begin{eqnarray}
    x_q[k,l] = \begin{cases} 
        \sqrt{\frac{E_{d,q}}{|\mathcal{D}^q|}} x_{d,q}[k,l] &\quad \text{if } (k,l) \in \mathcal{D}^q \\
        \sqrt{E_{p,q}} &\quad \text{if } (k,l) = (k_{p,q}, l_{p,q}) \\
        0 &\quad \text{otherwise},
    \end{cases}
\end{eqnarray}
where ${\mathcal D}_q$ is the set of DD bins/locations which carry data/information symbols and \( |\mathcal{D}^{q}| \) denotes the number of information symbols. Further, \( x_{d,q}[k,l] \) is the information symbol (e.g., $4$-QAM) transmitted on the \( (k,l) \)-th DD bin (i.e., carried by the quasi-periodic DD pulse located at $(k/B_q, l/T_q)$). We assume \( \mathbb{E}\left[|x_{d,q}[k,l]|^2\right] = 1 \). Therefore, the total transmitted average energy of all data symbols is $E_{d,q}$.
The  energy of the pilot symbol is $E_{p,q}$ and therefore the pilot-to-data-power ratio (PDR) is \( \frac{E_{p,q}}{E_{d,q}} \). 

With unit energy transmit and receive filter and normalized channel path gains in (\ref{wirelessch}) (i.e., ${\mathbb E}\left[ \sum\limits_{i=1}^{P_q} \vert h_{i,q} \vert^2 \right] = 1$), the received average energy of data symbols
is $E_{d,q}$ and therefore the average received data power is $E_{d,q}/T_q$. The noise power in bandwidth $B_q$ is \( N_0 B_q \), and therefore the signal-to-noise ratio (SNR) at the receiver is given by
\begin{eqnarray}
    \rho_q \Define \frac{E_{d,q}}{N_0 B_q T_q}.
\end{eqnarray}
At the receiver, the effective discrete channel taps \( h_{\mbox{\scriptsize{eff}},q,q}[k', l'] \) are estimated within the pilot region \( \mathcal{P}^q \) (i.e., $(k',l') \in \mathcal{P}^q$) using the estimation method described in equation $(30)$ of \cite{ZAKOTFS2}. The symbols received in the data and guard regions \( (\mathcal{D}^q \cup \mathcal{G}^q) \) are then used to detect the transmitted information symbols \( x_{d,q}[k,l] \) through joint DD domain equalization (see the matrix vector formulation of the Zak-OTFS I/O relation in \cite{ZAKOTFS2}).

    \section{Numerical Results and Discussions}
\label{numsec}
    \begin{table}[h!]
		\caption{{Simulation parameters for multiple UTs}}
		\label{parameters}
		\centering
		\begin{tabular}
			{ | c|| c| c| c| c| }
			\hline
			\multicolumn{5}{|c|}{Simulation parameters} \\
			\hline
			Parameters & UT-1 & UT-2 & UT-3 & UT-4\\
			\hline
            Doppler Period (KHz) & 15 & 15 & 30 & 30\\
			\hline
            No. of Delay bins ($M_s$) & 24 & 24 & 12 & 24\\
			\hline
            No. of Doppler bins ($N_s$) & 15 & 30 & 30 & 15\\
			\hline
            Bandwidth (KHz) & 360 & 360 & 360 & 720\\
			\hline
            Time duration (msec) & 1 & 2 & 1 & 0.5\\
			\hline
		\end{tabular}
	\end{table} 

\begin{table}[!t]
    \centering
    \caption{Power-delay profile of Veh-A channel model}
    \vspace{-2mm}
    \begin{tabular}{|c|c|c|c|c|c|c|}
         \hline
         Path index $i$ & 1 & 2 & 3 & 4 & 5 & 6 \\
         \hline
         Delay $\tau_{i,q} (\mu s)$ & 0 & 0.31 & 0.71 & 1.09 & 1.73 & 2.51 \\
         \hline
         Relative power (dB) & 0 & -1 & -9 & -10 & -15 & -20 \\
         \hline
    \end{tabular}
    \label{tab:veh_a}
    \vspace{-3mm}
\end{table}

In this section, we present the performance results from simulation
studies carried out for the proposed Zak-OTFS-MUL system with
$U=4$ UTs which are allocated TF resource as illustrated in Fig.~\ref{fig:MA_example} with their respective system parameters (time duration, bandwidth and Doppler period) listed in Table \ref{parameters}.
The uplink physical channel between each UT and the BS follows the vehicular-A channel model whose power-delay profile is listed in Table \ref{tab:veh_a} \cite{EVAITU}. The random channel realization for each UT is different.
In Table \ref{tab:veh_a}, for the $q$-th UT, the relative power of the $i$-th channel path w.r.t. the first path is ${\mathbb E}\left[ \vert h_{i,q} \vert^2 \right]/{\mathbb E}\left[ \vert h_{1,q} \vert^2 \right]$. The Doppler shift of the $i$-th path is given by $\nu_{i,q} = \nu_{max} \cos(\theta_{i,q})$, where $\theta_{i,q}$, $i=1,2,\cdots,6$, $q=1,2,3,4$ are i.i.d. uniformly distributed in $[0 \,,\, 2 \pi)$. We consider the sinc and root raised cosine (RRC) pulse shaping filters. The sinc pulse shaping filter for the $q$-th UT is given by

{\vspace{-4mm}
\small
\begin{eqnarray}
    w_{tx,q}(\tau, \nu) & \hspace{-2.5mm} = & \hspace{-2.5mm} \sqrt{B_q T_q}\,\text{sinc}(B_q\tau) \,  \text{sinc}(T_q \nu) \, e^{j 2 \pi (\nu_q \tau - \nu \tau_q)}.
\end{eqnarray}\normalsize}The RRC pulse shaping filter decays faster than the sinc pulse and therefore leaks less energy outside its main lobe due to which it experiences lesser DD domain aliasing and allows for operation at higher Doppler spread channels, but at the cost of time and bandwidth expansion \cite{ZAKOTFS2, otfsbook}. For our MA system, the RRC transmit pulse shaping filter for the $q$-th UT is given by

{\vspace{-4mm}
\small
\begin{eqnarray}
\label{rrcpulse_eqn1}
w_{tx,q}(\tau,\nu) & \hspace{-2.5mm} = &  \hspace{-2.5mm} \sqrt{B_q T_q} \, rrc_{_{\beta_{\tau,q}}}( B_q \tau ) \,  rrc_{_{\beta_{\nu,q}}}( T_q \nu ) \, e^{j 2 \pi (\nu_q \tau - \nu \tau_q)}, \nonumber \\
rrc_{_{\beta}}(x) &  = &  \frac{\sin(\pi x (1 - \beta)) + 4 \beta x \cos(\pi x (1 + \beta))}{\pi x \left( 1 - (4 \beta x)^2 \right)},
\end{eqnarray}\normalsize}where $\beta_{\tau,q}$ and $\beta_{\nu,q}$ are the roll-off factors. Due to time and bandwidth expansion, for the $q$-th UT, the actual time-duration and bandwidth of the Zak-OTFS frame is $(1 + \beta_{\nu,q})T_q$ and $(1 + \beta_{\tau,q})B_q$ respectively.
For simulations, we consider $\beta_{\tau,q} = \beta_{\nu,q} = 0.1$ for all UTs. The matched filter at the receiver is related to the transmit pulse shaping filter and is given by
$w_{rx,q}(\tau, \nu) = w_{tx}^*(-\tau, -\nu) \, e^{j 2 \pi \nu \tau}$ \cite{Hanly23}. Note that all UTs use the same type of pulse shaping, i.e. either they all use sinc or all use RRC. Also, in Fig.~\ref{fig:modelfree}, for all UTs $a_1 = 2$, $a_2 = 1$, $g_1 = 3$ and $g_2 = 2$.
    \begin{figure}
            \vspace{-0.8cm}
			\includegraphics[width=1\linewidth, height=0.9\linewidth]{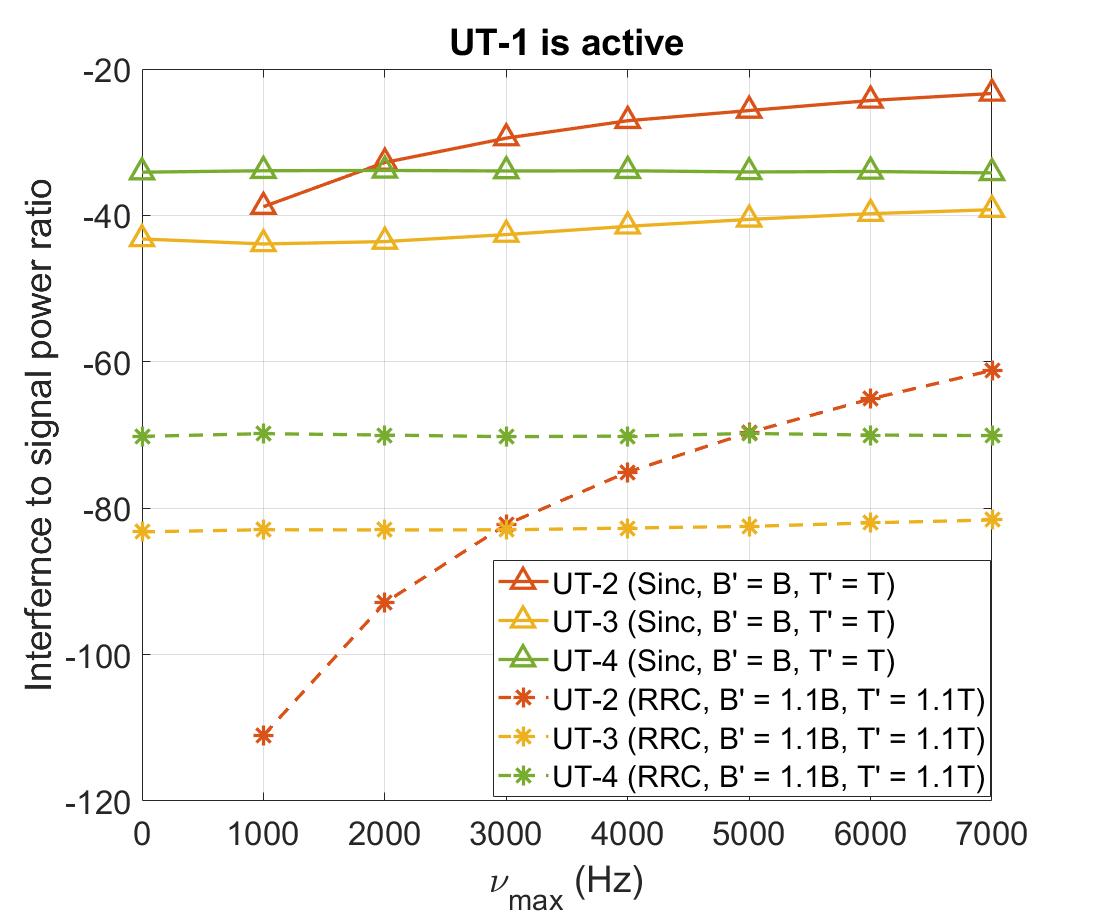}
            \vspace{-2mm}
			\caption{Leakage ratio $\frac{I_{q,s}}{S_{s,s}}$ (see (\ref{eqnratio45})) vs. $\nu_{max}$ when only UT-1 is transmitting}
			\label{MUI_nu_maxUT1}
	\end{figure}
        \begin{figure}
            \vspace{-0.8cm}
			\includegraphics[width=1\linewidth, height=0.9\linewidth]{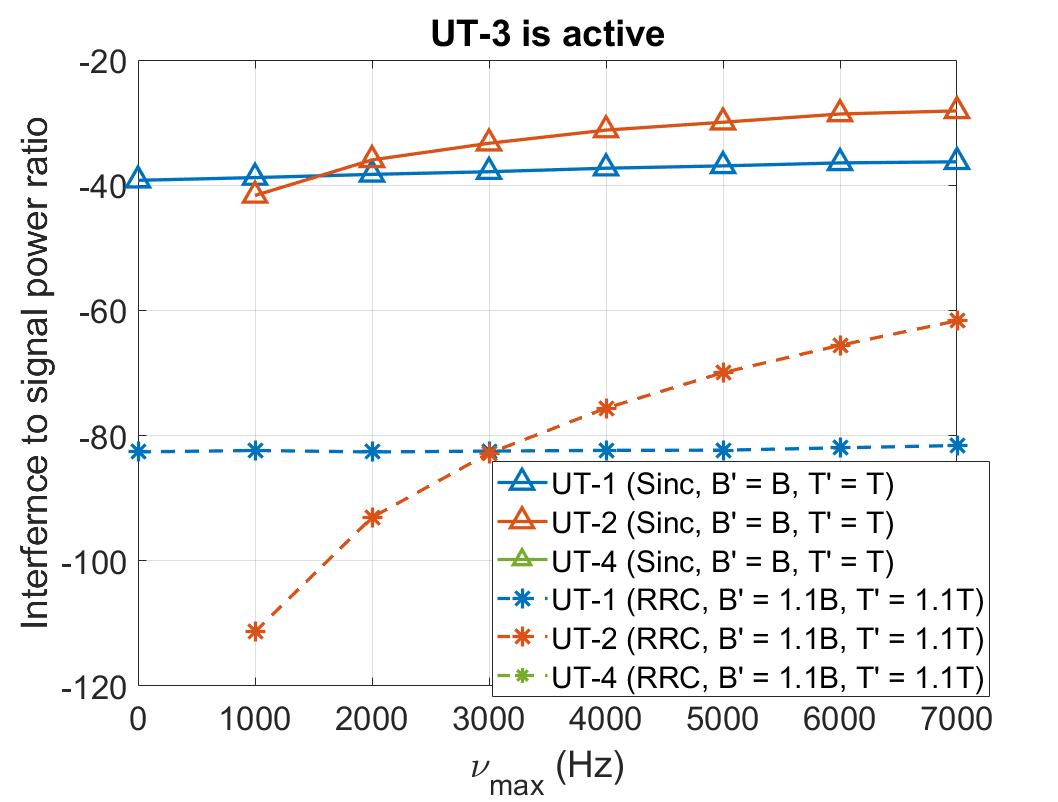}
            \vspace{-2mm}
			\caption{Leakage ratio $\frac{I_{q,s}}{S_{s,s}}$ (see (\ref{eqnratio45})) vs. $\nu_{max}$ when only UT-3 is transmitting}
			\label{MUI_nu_maxUT3}
	\end{figure}

    In Fig.~\ref{MUI_nu_maxUT1} we study the average leakage to useful signal power ratio $\frac{I_{q,s}}{S_{s,s}}$ in (\ref{eqnratio45}) as a function of increasing $\nu_{max}$ for $s=1$ (i.e., only UT-1 is transmitting) and we measure the leaked energy to the matched filter output for the three remaining UTs $q=2,3,4$. It is observed that the energy leakage is higher with sinc filter than with RRC since the sinc pulse decays slowly as compared to RRC. This reflects the sensitivity of the interference to signal ratio to the choice of pulse-shaping filter at the user transmitter and matched-filter at the BS receiver.
    
    In Fig.~\ref{MUI_nu_maxUT1}, the leakage for UT-2 increases monotonically with increasing $\nu_{max}$ since they share common time resource and adjacent frequency resource (see Fig.~\ref{fig:MA_example}). With increasing $\nu_{max}$, due to Doppler shift the energy transmitted by UT-1 in frequency resources adjacent/bordering those allocated to UT-2, leaks into UT-2 along the frequency domain. With increasing $\nu_{max}$ the Doppler shift also increases thereby increasing the leakage monotonically. Interestingly, with RRC filters, this leakage to UT-4 is less than $-60$ dB even for an extremely high $\nu_{max} = 7$ KHz (i.e., Doppler spread of $14$ KHz). 

    Further, the leakage to UT-3 and UT-4 is almost invariant of increasing $\nu_{max}$, since the TF resources allocated to UT-1, UT-3 and UT-4 do not overlap in time and therefore even if they communicate on the same/adjacent frequency resources they do so on different time intervals (see TF allocation in Fig.~\ref{fig:MA_example}). 
    The leakage to UT-3 and UT-4 is however not zero due to leakage along time domain as they occupy time resources adjacent to UT-1. The leakage along time-domain is due to the delay spread of the effective discrete DD channel filter of UT-1.
    
    

    In Fig.~\ref{MUI_nu_maxUT3} we plot the leakage ratio when only UT-3 is transmitting. The leakage to UT-2 increases monotonically with increasing $\nu_{max}$ since UT-2 and UT-3 are allocated adjacent frequency resources and share common time resource. Since the TF allocation of UT-3 and UT-4 is neither adjacent along time nor along frequency (see Fig.~\ref{fig:MA_example}), the leakage from UT-3 to UT-4 is almost zero (leakage ratio is less than $-120$ dB and is therefore not visible in the figure). Since UT-1 is allocated adjacent non-overlapping time resource it experiences some leakage which is almost invariant of $\nu_{max}$. 
    \begin{figure}
    \hspace{-4mm}
			\includegraphics[width=9.8cm, height=6.7cm]{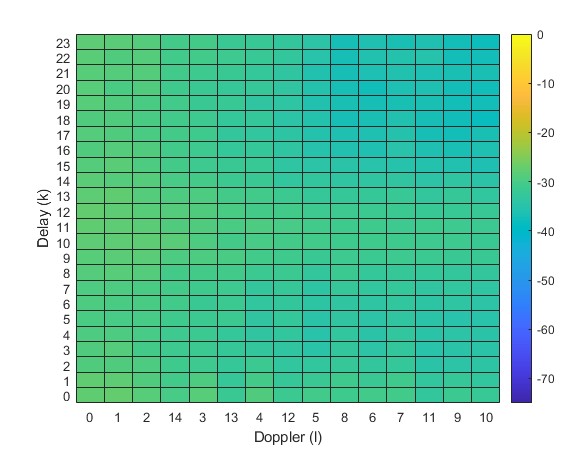}
			\caption{Heatmap of the ratio of MUI received to the useful signal power, for each DD carrier at the match-filtered output of UT-1, with all users active and transmitting the same average power. This ratio for the $(k,l)$-th carrier of UT-1 is $\frac{   {\mathbb E}\left[ \sum\limits_{q=2}^U \int\limits_{0}^{\tau_{p,q}}\int\limits_{0}^{\nu_{p,q}} \left\vert {\Tilde h}_{1,q}^{k,l}(\tau, \nu) \right\vert^2 \, d\tau \, d\nu \right] }{   {\mathbb E}\left[ \int\limits_{0}^{\tau_{p,1}}\int\limits_{0}^{\nu_{p,1}} \left\vert {\Tilde h}_{1,1}^{k,l}(\tau, \nu) \right\vert^2 \, d\tau \, d\nu \right]}$. Data-only frames are transmitted by all users, i.e., there is no pilot/guard region. $\nu_{max} = 1$ KHz (Doppler spread is $2$ KHz), sinc pulse shaping filter. Other parameters and TF allocation is the same as in Fig.~\ref{MUI_nu_maxUT1}. Note that the ratio is almost uniform in the DD domain.}
			\label{heatmapMUI}
	\end{figure}
    
        In both Fig.~\ref{MUI_nu_maxUT1} and Fig.~\ref{MUI_nu_maxUT3} it is noted that with sinc pulse-shaping and matched-filter, the interference to signal ratio is lower than $-30$ dB even for a very high Doppler spread of $6$ KHz, and without any guard TF resource between adjacent TF allocations. Since the SNR is usually below $30$ dB, \emph{interference is therefore dominated by AWGN}. 
        Also, although the interference energy in the TF domain is localized along the boundary between TF allocations, in the DD domain the interference energy is spread almost uniformly across all DD carriers (see Fig.~\ref{heatmapMUI}). In other words, a single DD carrier is not adversely affected by interference.\footnote{\footnotesize{The discrete DD domain match-filtered output of the $s$-th user is simply the projection of the received TD signal onto the TD realization of the carriers of the $s$-th user \cite{otfsbook}. This is also equivalent to the projection of the FD realization of the received signal onto the FD realization of the carriers. 
        The TD realization of a carrier of the $s$-th user consists of $N_s$ TD pulses within the allocated time interval $[\tau_s-T_s/2, \tau_s+T_s/2]$ and spaced regularly apart by the delay period $\tau_{p,s} = M_s/B_s$. The FD realization of a carrier consists of $M_s$ FD pulses within the allocated FD interval $[\nu_s-B_s/2, \nu_s+B_s/2]$ and spaced regularly apart by the Doppler period $\nu_{p,s} = N_s/T_s$. The TD and FD realization of each DD carrier is therefore spread out uniformly across the allocated TF resource. Hence the interference signal localized at the boundary between adjacent TF allocations has almost similar projections with all the $M_s N_s$ DD carriers, i.e., the interference energy is distributed almost uniformly in the DD domain.}} {These observations imply single-user performance in multiuser uplink}. This is indeed the case as we see next.

\begin{figure}
    \hspace{-4mm}
			\includegraphics[width=9.8cm, height=6.7cm]{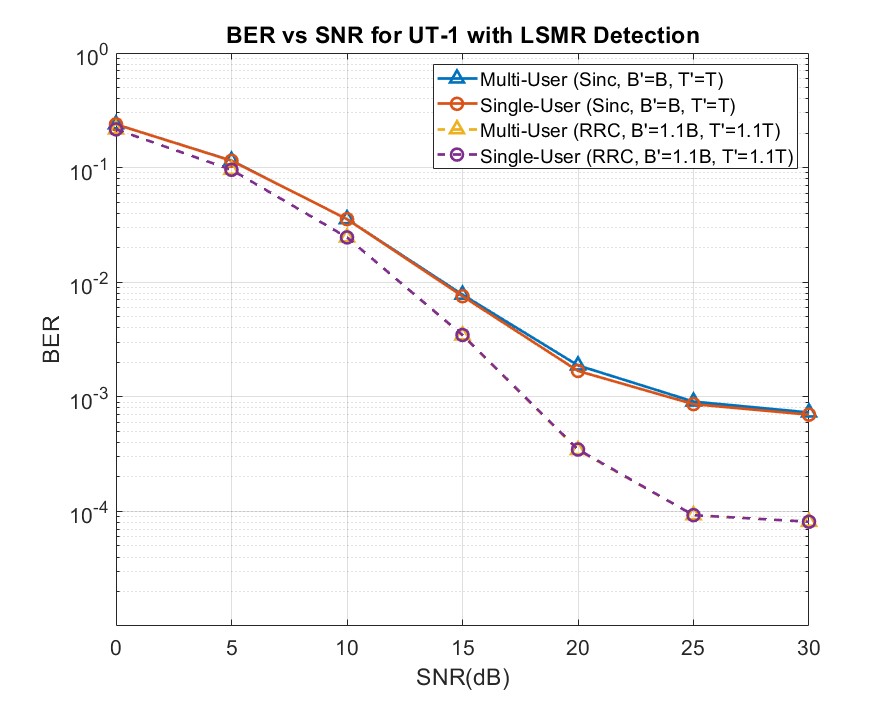}
			\caption{BER vs. SNR for UT-1 for proposed multiuser allocation and single-user with embedded pilot-data Zak-OTFS frame. Fixed PDR$=0$ dB and $\nu_{max} = 815$ Hz for all UTs. LSMR equalization.}
			\label{bervssnr}
	\end{figure}

    Next in Fig.~\ref{bervssnr} we plot the bit error rate (BER) for the detection of uncoded $4$-QAM symbols transmitted by UT-1 in the presence of transmission by all other UTs (all UTs transmit embedded pilot-data Zak-OTFS frame with uncoded $4$-QAM information symbols). From the match-filtered output for UT-1, the BS first estimates the effective discrete DD channel filter taps which is then used to equalize and detect the transmitted symbols using the Least Squares Minimum Residual (LSMR) equalizer proposed for single-user MC-OTFS in \cite{LSMR}. The PDR is $0$ dB and $\nu_{max} = 815$ KHz for all UTs.
    We also plot the BER for a single-user system where only UT-1 is transmitting.
    It is observed that for both the sinc and the RRC filters, the BER performance for the multiuser system is same as that for the single-user system. 
    Also, for both the single-user and multiuser system, the
    BER performance floors at high SNR greater than $25$ dB, which is primarily due to the residual error in the estimated taps of the effective discrete DD channel filter.

\begin{figure}
    \hspace{-4mm}
			\includegraphics[width=9.8cm, height=6.7cm]{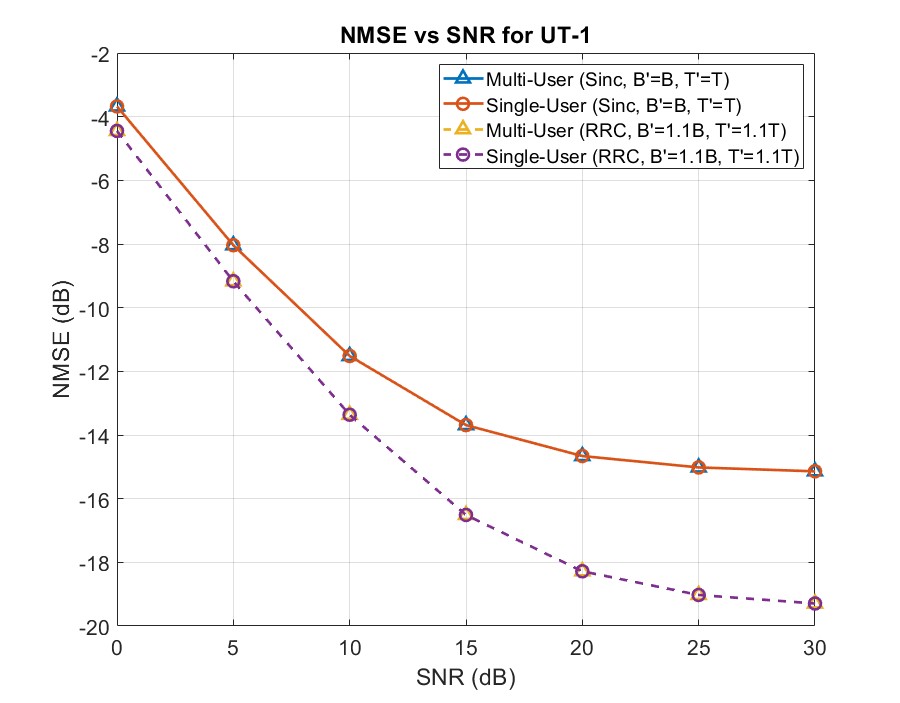}
			\caption{NMSE vs. SNR for UT-1 for proposed multiuser allocation and single-user with embedded pilot-data Zak-OTFS frame. Fixed PDR$=0$ dB and $\nu_{max} = 815$ Hz for all UTs.}
			\label{nmsevssnr}
	\end{figure}

    To see this, in Fig.~\ref{nmsevssnr} we plot the normalized mean squared error (NMSE) of the estimated taps of the effective channel filter as a function of increasing SNR.
    Let ${\widehat h}_{\mbox{\scriptsize{eff}},q,q}[k,l]$ denote the estimated taps of of the effective channel filter for the $q$-th UT. Then, the NMSE for the $q$-th UT is given by $\frac{{\mathbb E}\left[ \sum\limits_{(k,l) \in {\mathcal S}_q} \left\vert {\widehat h}_{\mbox{\scriptsize{eff}},q,q}[k,l] - { h}_{\mbox{\scriptsize{eff}},q,q}[k,l] \right\vert^2 \right]}{ {\mathbb E}\left[ \sum\limits_{(k,l) \in {\mathcal S}_q} \left\vert  { h}_{\mbox{\scriptsize{eff}},q,q}[k,l] \right\vert^2\right]}$, where ${\mathcal S}_q$ denotes the DD domain support set of ${ h}_{\mbox{\scriptsize{eff}},q,q}[k,l]$. Indeed for SNR greater than $25$ dB the NMSE performance floors. Also, the NMSE performance for both single- and multiuser systems is the same which again shows that there is negligible MUI.

        \begin{figure}
    \hspace{-4mm}
			\includegraphics[width=9.8cm, height=6.7cm]{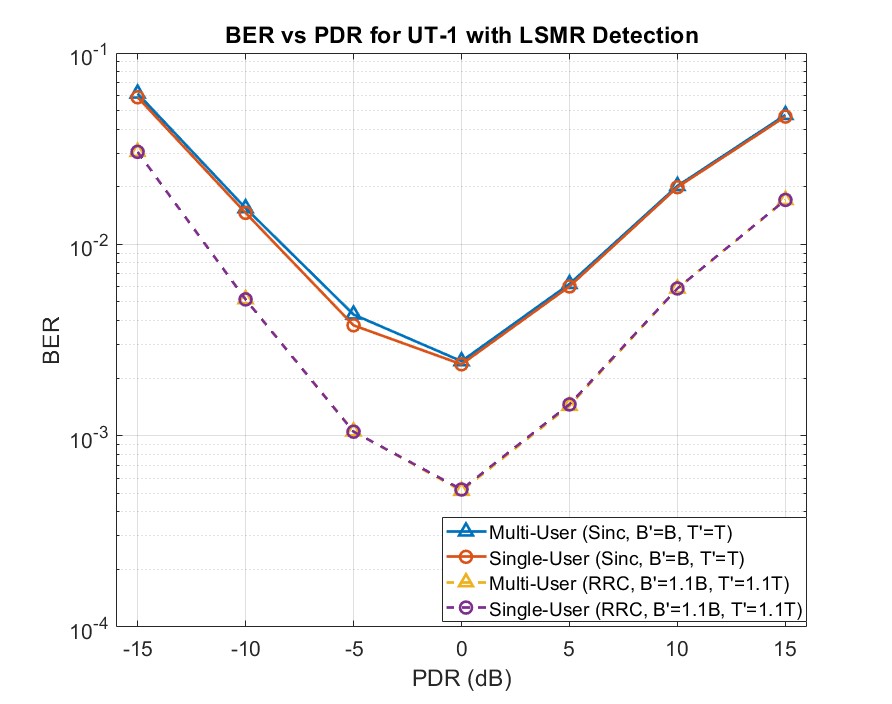}
			\caption{BER vs. PDR for UT-1 for proposed multiuser allocation and single-user with embedded pilot-data Zak-OTFS frame. Fixed SNR $\rho_q=20$ dB and $\nu_{max} = 815$ Hz for all UTs. LSMR equalization.}
			\label{bervspdr}
	\end{figure}
    
Next, in Fig.~\ref{bervspdr} we investigate the dependence of the BER on the PDR for a fixed SNR of $20$ dB and $\nu_{max} = 815$ Hz for all UTs. Again, the single- and multiuser performance are identical, and both exhibit a characteristic ``U" shaped curve with optimal BER at PDR of $0$ dB. The ``U" shape is because, with increasing PDR, the pilot becomes stronger which increases the channel estimation accuracy (i.e., reduces NMSE) resulting in BER improvement. However, when PDR exceeds $0$ dB, leakage of the received pilot energy to the information carriers in the data region becomes more dominant than noise which results in degradation in BER performance.

\begin{figure}
    \hspace{-4mm}
			\includegraphics[width=9.8cm, height=6.7cm]{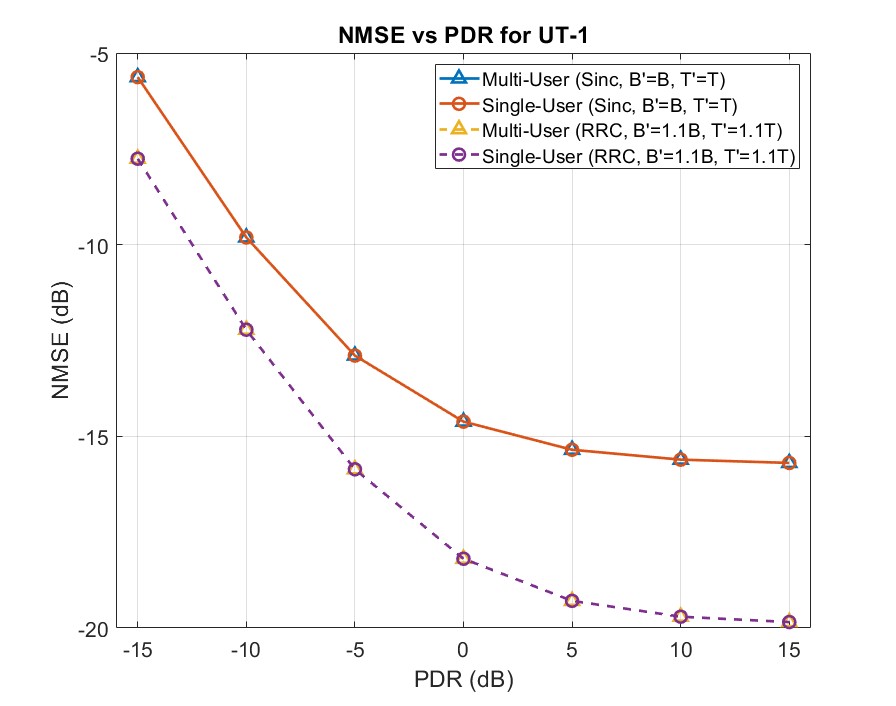}
			\caption{NMSE vs. PDR for UT-1 for proposed multiuser allocation and single-user with embedded pilot-data Zak-OTFS frame. Fixed SNR $\rho_q=20$ dB and $\nu_{max} = 815$ Hz for all UTs.}
			\label{nmsevspdr}
	\end{figure}

In Fig.~\ref{nmsevspdr} we plot NMSE as a function of increasing PDR for a fixed SNR of $20$ dB and fixed $\nu_{max} = 815$ Hz for all UTs. Indeed, with increasing PDR, the pilot becomes stronger and the accuracy of the estimated effective channel filter improves thereby reducing NMSE monotonically. However, when PDR exceeds $0$ dB, further NMSE improvement saturates since the NMSE is limited by the small energy of the effective channel filter which leaks outside the pilot region ${\mathcal P}^q$ (see Fig.~\ref{fig:modelfree}). Also, note that the NMSE performance for both single and multiuser scenarios is the same.

\begin{figure}
    \hspace{-4mm}
			\includegraphics[width=9.8cm, height=6.7cm]{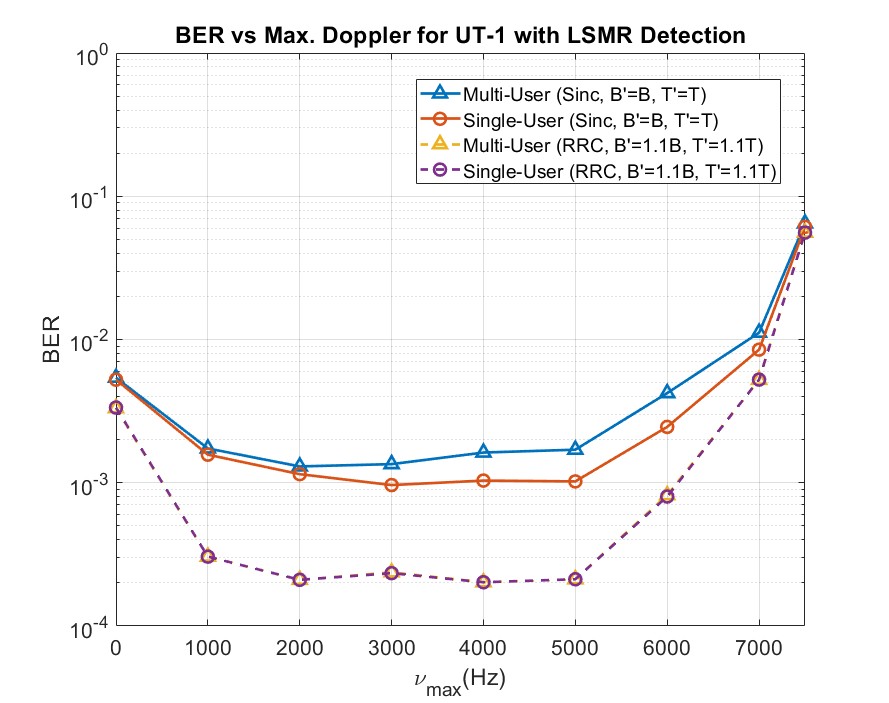}
			\caption{BER vs. $\nu_{max}$ for UT-1 for proposed multiuser allocation and single-user with embedded pilot-data Zak-OTFS frame. Fixed SNR $\rho_q=20$ dB and PDR $=0$ dB for all UTs. LSMR equalization.}
			\label{bervsnumax}
	\end{figure}

In Fig.~\ref{bervsnumax} we plot BER vs. $\nu_{max}$ for a fixed SNR of $20$ dB and PDR of $0$ dB for all UTs. For UT-1, the crystallization condition is satisfied as long as the Doppler spread $(2 \nu_{max} + 1/T_1)$ of the effective channel filter ${ h}_{\mbox{\scriptsize{eff}},1,1}(\tau, \nu)$
is less than the Doppler period $\nu_{p,1} = 15$ KHz (see Table \ref{parameters}). Since the time duration of the TF resource allocated to UT1 i.e. $T_1 = 1$ ms, this condition is met when $\nu_{max} < \frac{(\nu_{p,1} - 1/T_1)}{2} = 7$ KHz. Indeed, in Fig.~\ref{bervsnumax}, BER is less than $10^{-2}$ for $\nu_{max} < 7$ KHz and degrades beyond that. Again, the BER performance for single- and multiuser scenarios is almost identical. \emph{Low MUI therefore allows us to achieve BER performance robust to very high Doppler spreads even in a multiuser uplink channel}.

            \begin{figure}
    \hspace{-4mm}
			\includegraphics[width=9.8cm, height=6.7cm]{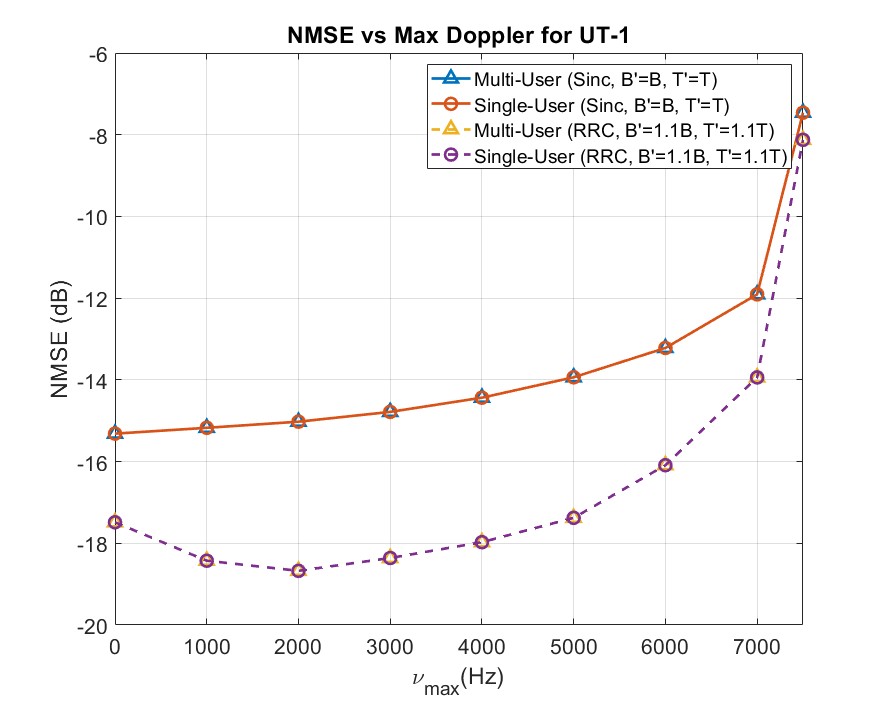}
			\caption{NMSE vs. $\nu_{max}$ for UT-1 for proposed multiuser allocation and single-user with embedded pilot-data Zak-OTFS frame. Fixed SNR $\rho_q=20$ dB and PDR $=0$ dB for all UTs.}
			\label{nmsevsnumax}
	\end{figure}

In Fig.~\ref{nmsevsnumax} we plot NMSE vs. $\nu_{max}$ for a fixed SNR of $20$ dB and PDR of $0$ dB for all UTs. It is observed that the NMSE performance is good for $\nu_{max} < 7$ KHz and degrades severely beyond that. This is because, for $\nu_{max} \geq 7$ KHz, there is Doppler domain aliasing between ${\mathcal A}^1$ (see (\ref{eqn52523})) and the support sets of the terms corresponding to $(n,m) \ne (0,0)$ in (\ref{eqn52523}), which degrades the accuracy of the estimated effective discrete DD channel filter. Again, the NMSE performance for both single- and multiuser scenarios is identical.

    \section{Conclusion}
In this paper we described a new method of shaping a transmitted Zak-OTFS pulse in the DD domain that enables non-overlapping allocation of TF resources. We demonstrated that it is possible to choose the delay and Doppler periods that define Zak-OTFS modulation to match the channel characteristics of an individual user, and to do so independently of the choices made for other users.
The base station receives a superposition of uplink signals and applies individual matched filters to obtain the data specific to individual users. We described the method of designing pulse-shaping transmit filter and the corresponding matched filter that control interference between different users at the base station receiver.
We demonstrated single-user performance in a multiuser Zak-OTFS uplink system without needing to provision guard bands between TF resources allocated to different users. These performance results show that the benefits of a predictable Zak-OTFS waveform can be realized within an architecture for uplink communication that enables users with different characteristics to share spectrum.

	\appendices

        \section{Proof of (\ref{eqn8})}
    \label{appprfb}
     It suffices to show that starting with (\ref{eqn8}) relating $a(t)$ and $b(t)$ and taking the Zak transform of both sides, we get the relation between the DD representation of these two signals, i.e., $a_{dd}(\tau, \nu) = w(\tau, \nu) \, *_{\sigma} \, b_{dd}(\tau, \nu)$. Taking Zak transform of both sides of (\ref{eqn8}) gives

     {\small
     \vspace{-4mm}
     \begin{eqnarray}
         a_{dd}(\tau, \nu) & \mya & \sqrt{\tau_{p,s}} \sum\limits_{k \in {\mathbb Z}} a(\tau + k \tau_{p,s}) \, e^{-j 2 \pi k \nu \tau_{p,s}} \nonumber \\
         & & \hspace{-17mm} \myb \sqrt{\tau_{p,s}} \sum\limits_{k \in {\mathbb Z}} \iint {\Big [} w(\tau', \nu') b(\tau + k \tau_{p,s} - \tau') e^{j 2 \pi \nu' (\tau - \tau' + k \tau_{p,s})} \nonumber \\
         & & \hspace{20mm} \, e^{-j 2 \pi k \nu \tau_{p,s}} {\Big ]} \, d\tau' \, d\nu' \nonumber \\
         & & \hspace{-17mm} \myc \iint w(\tau', \nu') \underbrace{{\Big [} \sqrt{\tau_{p,s}} \sum\limits_{k \in {\mathbb Z}} b(\tau - \tau' + k \tau_{p,s}) \, e^{-j 2 \pi (\nu - \nu') k \tau_{p,s}}   {\Big ]}}_{= b_{dd}(\tau - \tau', \nu - \nu')}  \nonumber \\
         & & \hspace{20mm} e^{j 2 \pi \nu' (\tau - \tau')} \, d\tau' \, d\nu' \nonumber \\
         & & \hspace{-17mm} = \iint w(\tau', \nu') b_{dd}(\tau - \tau', \nu - \nu') \, e^{j 2 \pi \nu' (\tau - \tau')} \, d\tau' \, d\nu' \nonumber \\
         & & \hspace{-17mm} \myd \, w(\tau, \nu) \, *_{\sigma} b_{dd}(\tau, \nu),
     \end{eqnarray}\normalsize}where step (a) follows from the definition of the Zak transform in (\ref{eqn7}). Step (b) follows from substituting the expression of $a(t)$ in (\ref{eqn8}) into the R.H.S. in step (a). Step (c) follows from swapping the order of summation and integration in step (b). Step (d) following from the definition of twisted convolution in (\ref{eqn2}).
     
    \section{Derivation of step-(a) in (\ref{eqn1213})}
	\label{appendix_td_pulsone}
    The TD representation of the carrier for the $(k,l)$-th information symbol, i.e., $\phi^{k,l}_{s}(t)$ is given by
    {\small
    \begin{eqnarray}
    \label{A1_1}
        \phi^{k,l}_{s} (t) &\hspace{-1mm}=&\hspace{-1mm}  \mathcal{Z}^{-1}_t\left(  \phi_{dd,s}^{k,l}(\tau, \nu) \right) \nonumber \\
        & & \hspace{-17mm} \mya  \sqrt{\tau_{p,s}} \int_0^{\nu_{p,s}} \phi_{dd,s}^{k,l}(t, \nu) d\nu \nonumber \\
         & & \hspace{-17mm} \myb \sum_{n,m \in {\mathbb Z}}  \hspace{-1mm} e^{j 2 \pi \frac{nl}{N_s}} \delta\hspace{-1mm}\left(\hspace{-1mm} t - n\tau_{p,s} - \frac{k \tau_{p,s}}{M_s} \hspace{-1mm}\right)  \int\limits_{0}^{\nu_{p,s}}  \hspace{-1mm} \delta\hspace{-1mm}\left(\hspace{-1mm}\nu - m\nu_{p,s} - \frac{l \nu_{p,s}}{N_s}\hspace{-1mm}\right) \, d\nu \nonumber \\
        &\hspace{-17mm}\myc &\hspace{-1mm} \sqrt{\tau_{p,s}} \sum_{n \in \mathbb{Z}} e^{j 2 \pi \frac{nl}{N_s}} \delta \left( t - n \tau_{p,s} -\frac{k\tau_{p,s}}{M_s}\right). 
    \end{eqnarray}}
    where step (a) follows using the definition of inverse Zak Transform and step (b) follows from using the expression of $\phi_{dd,s}^{k,l}(\tau, \nu)$ in  (\ref{basis_sig}). Step (c) follows from the fact that for all $l=0,1,\cdots, N_s -1 $, the integral $\int\limits_{0}^{\nu_{p,s}}  \hspace{-1mm} \delta\hspace{-1mm}\left(\hspace{-1mm}\nu - m\nu_{p,s} - \frac{l \nu_{p,s}}{N_s}\hspace{-1mm}\right) \, d\nu$ in the R.H.S. is zero for all $m \ne 0$ and is one for $m=0$.

    \section{Derivation of (\ref{eqn15})}
	\label{appendix_eqn16}
Substituting $w_{tx,s} = w_{B_s}(\tau) \, w_{T_s}(\nu)$ in the RHS of (\ref{eqn14}) gives

{\small
\vspace{-4mm}
\begin{eqnarray}
    \psi^{k,l}_s(t) & \hspace{-3mm} = & \hspace{-3mm} \iint w_{B_s}(\tau) \, w_{T_s}(\nu)  \, \phi^{k,l}_s(t - \tau) \, e^{j 2 \pi \nu (t - \tau)} \, d\tau \, d\nu \nonumber \\
& & \hspace{-14mm} = \int w_{B_s}(\tau) \phi_s^{k,l}(t - \tau) \underbrace{\left[ \int w_{T_s}(\nu) \, e^{j 2 \pi \nu (t - \tau)} \, d\nu \right]}_{= W_{T_s}(t - \tau)} d\tau,
\end{eqnarray}\normalsize}where $W_{T_s}(t) = \int w_{T_s}(\nu) \, e^{j 2 \pi \nu t} \, d\nu$ is the inverse Fourier transform of $w_{T_s}(\nu)$. Therefore

{\small
\vspace{-4mm}
\begin{eqnarray}
    \psi^{k,l}_s(t) & \hspace{-3mm} = & \hspace{-3mm} \int w_{B_s}(\tau) \phi_s^{k,l}(t - \tau)  W_{T_s}(t - \tau) \, d\tau \nonumber \\
& \hspace{-3mm} = & \hspace{-3mm} w_{B_s}(t) \, \star \, \left[ W_{T_s}(t) \, \phi_s^{k,l}(t) \right].
\end{eqnarray}\normalsize}

    \section{Proof of Theorem \ref{theorem2}}
	\label{appendix_theorem_2}
    We factorize ${\Tilde w}_{tx,s}(\tau, \nu)$ as
\begin{eqnarray}
\label{eqn19746}
{\Tilde w}_{tx,s}(\tau, \nu) & = & {\Tilde w}_{T_s}(\nu) \, {\Tilde w}_{B_s}(\tau), \nonumber \\
{\Tilde w}_{T_s}(\nu) & \Define & w_{T_s}(\nu) \, e^{-j 2 \pi \tau_s \nu}, \nonumber \\
{\Tilde w}_{B_s}(\tau) & \Define & w_{B_s}(\tau) \, e^{j 2 \pi \nu_s \tau}.
\end{eqnarray}Therefore, from (\ref{eqn15}) it follows that with this new pulse shaping filter, the carrier waveform for the $(k,l)$-th information symbol is given by
    \begin{eqnarray}
\label{wytce}
         {\Tilde \psi}^{k,l}_s(t) & \hspace{-2mm} = & \hspace{-2mm} {\Tilde w}_{B_s}(t) \, \star \, \left[ {\Tilde W}_{T_s}(t) \,  \phi^{k,l}_s(t) \right]
    \end{eqnarray}where ${\Tilde W}_{T_s}(t) = \int {\Tilde w}_{T_s}(\nu) \, e^{j 2 \pi \nu t} \, d\nu$. Substituting (\ref{eqn19746}) in (\ref{wytce}) we get
    \begin{eqnarray}
\label{wytce3}
         {\Tilde \psi}^{k,l}_s(t) & \hspace{-2mm} = & \hspace{-2mm} ({w}_{B_s}(t) \, e^{j 2 \pi \nu_s t})  \, \star \, \left[ {W}_{T_s}(t - \tau_s) \,  \phi^{k,l}_s(t) \right],
    \end{eqnarray}since
\begin{eqnarray}
{\Tilde W}_{T_s}(t) & \hspace{-2mm} = & \hspace{-2mm} \int {\Tilde w}_{T_s}(\nu) \, e^{j 2 \pi \nu t} \, d\nu \nonumber \\
& \hspace{-2mm} = &  \hspace{-2mm} \int w_{T_s}(\nu) \, e^{-j 2 \pi \tau_s \nu } \,  e^{j 2 \pi \nu t} \, d\nu  \nonumber \\
& \hspace{-2mm} = &  \hspace{-2mm}  \int w_{T_s}(\nu) \,  e^{j 2 \pi \nu (t - \tau_s)} \, d\nu  \, = \, W_{T_s}(t - \tau_s),
\end{eqnarray}where the last step follows from the definition of $W_{T_s}(t)$ in (\ref{eqn15}). 

In the RHS in (\ref{wytce3}), the support interval of $W_{T_s}(t - \tau_s)$ is the support interval of $W_{T_s}(t)$ shifted by $\tau_s$, i.e., $\left[ \tau_s - \frac{T_s}{2} \,,\, \tau_s + \frac{T_s}{2} \right]$. Therefore, ${\Tilde \psi}^{k,l}_s(t) $ is approximately time limited to $\left[ \tau_s - \frac{T_s}{2} \,,\, \tau_s + \frac{T_s}{2} \right]$. The Fourier transform of ${\Tilde \psi}^{k,l}_s(t)$ in (\ref{wytce3}) is given by
\begin{eqnarray}
{\Tilde \Psi}^{k,l}_s(f) & \hspace{-2mm} = & \hspace{-2mm} W_{B_s}(f - \nu_s) \, \int \phi_{s}^{k,l}(f - \nu) \, w_{T_s}(\nu) \, e^{-j 2 \pi \tau_s \nu} \, d\nu. \nonumber \\
\end{eqnarray}The support interval of $W_{B_s}(f - \nu_s)$ is the support interval of $W_{B_s}(f)$ shifted by $\nu_s$, i.e., $\left[ \nu_s - \frac{B_s}{2} \,,\, \nu_s + \frac{B_s}{2} \right]$. Therefore,in the frequency domain ${\Tilde \psi}^{k,l}_s(t) $ is approximately limited to $\left[ \nu_s - \frac{B_s}{2} \,,\, \nu_s + \frac{B_s}{2} \right]$.

    \section{Proof of Theorem \ref{theorem4}}
	\label{appendix_theorem_4}
    Substituting (\ref{wirelessch}) into (\ref{effcontch}) gives (\ref{AE_1}) at the top of next page. Using the integral expression for twisted convolution results in (\ref{AE_2}) and further simplification of the integral gives (\ref{AE_4}). Further application of the remaining twisted convolution operation in  (\ref{AE_4}) and simplifying it gives final expression for $h_{\mbox{\scriptsize{eff}},q,s}(\tau, \nu)$ in (\ref{AE_6}).
    
    {
    \begin{figure*}
			{\vspace{-9mm} \small
				\begin{eqnarray}
					\label{AE_1}
	h_{\mbox{\scriptsize{eff}},q,s} (\tau, \nu) & \hspace{-1.5mm} = &  \hspace{-1.5mm} \left(  w^{*}_{B_q}(-\tau) w^{*}_{T_q}(-\nu) e^{j 2 \pi (\nu_q \tau- \nu\tau_q)} e^{j 2 \pi \nu \tau} \right)  *_{\sigma} \left(\sum_{i=1}^{P_s} h_{i,s} \delta(\tau - {\tau}_{i,s}) \delta(\nu - {\nu}_{i,s}) *_{\sigma} \, w_{B_s}(\tau) \, w_{T_s}(\nu) \, e^{j 2 \pi (\nu_s \tau - \nu \tau_s)} \right) \\
    \label{AE_2}
    &  &  \hspace{-23mm} = w^{*}_{tx,q}(-\tau, -\nu) e^{j 2 \pi \nu \tau} *_{\sigma} \hspace{-1mm} \int \hspace{-2mm} \int \hspace{-1mm} \sum_{i=1}^{P_s} \hspace{-1mm} h_{i,s} \delta(\tau' - {\tau}_{i,s}) \delta(\nu' - {\nu}_{i,s}) 
    w_{B_s}(\tau - \tau') \, w_{T_s}(\nu - \nu') \, e^{j 2 \pi (\nu_s (\tau - \tau') - (\nu - \nu') \tau_s)}  \, e^{j 2 \pi \nu' (\tau - \tau')} 
    d\tau'  d\nu' \\
    \label{AE_4}
    & & \hspace{-23mm} =   w^{*}_{B_q}(-\tau) w^{*}_{T_q}(-\nu) e^{j 2 \pi (\nu_q \tau- \nu\tau_q)} e^{j 2 \pi \nu \tau} *_{\sigma} \sum_{i=1}^{P_s} h_{i,s}  w_{B_s}(\tau-\tau_{i,s}) w_{T_s}(\nu-\nu_{i,s}) e^{j 2 \pi (\nu_s (\tau-\tau_{i,s})- (\nu-\nu_{i,s})\tau_s)} e^{j 2 \pi\nu_{i,s}(\tau-\tau_{i,s})} \\
    \label{AE_5}
     & & \hspace{-23mm} =  \sum_{i=1}^{P_s} h_{i,s} \int \hspace{-2mm} \int w^{*}_{B_q}(-\tau') w^{*}_{T_q}(-\nu') w_{B_s}(\tau-\tau'-\tau_{i,s}) w_{T_s}(\nu-\nu'-\nu_{i,s}) e^{j 2 \pi (\nu_q- \nu_s - \nu_{i,s})\tau'} e^{j 2 \pi \nu'(\tau - (\tau_q-\tau_s))} \nonumber \\
    && \hspace{80mm} \times e^{j 2 \pi \nu_s(\tau - \tau_{i,s})} e^{j 2 \pi \nu_{i,s}(\tau +\tau_s - \tau_{i,s})} e^{-j2\pi\nu\tau_s} d\tau' d\nu' \\
      & & \hspace{-23mm} =  \sum_{i=1}^{P_s} h_{i,s} e^{j 2 \pi \nu_s(\tau - \tau_{i,s})} e^{j 2 \pi \nu_{i,s}(\tau +\tau_s - \tau_{i,s})} e^{-j2\pi\nu\tau_s} \left(\int w^{*}_{B_q}(-\tau') w_{B_s}(\tau-\tau'-\tau_{i,s}) e^{j 2 \pi (\nu_q- \nu_s - \nu_{i,s})\tau'} d\tau'\right) \nonumber \\
    && \hspace{70mm} \times \left(\int  w^{*}_{T_q}(-\nu')  w_{T_s}(\nu-\nu'-\nu_{i,s})  e^{j 2 \pi \nu'(\tau - (\tau_q-\tau_s))} d\nu'\right) \\
     \label{AE_6}
      & & \hspace{-23mm} =  \sum_{i=1}^{P_s} h_{i,s} e^{j 2 \pi \nu_s(\tau - \tau_{i,s})} e^{j 2 \pi \nu_{i,s}(\tau +\tau_s - \tau_{i,s})} e^{-j2\pi\nu\tau_s}  \times \zeta_{q,s,i}(\tau) \times \eta_{q,s,i}(\tau, \nu)
				\end{eqnarray}
			}
			{
				\begin{eqnarray*}
					\hline
			\end{eqnarray*}}
				\normalsize
	\end{figure*}}

\end{document}